\documentclass[apj]{emulateapj}

\newcommand{\etal}{{\it et al.}}
\newcommand{\lsim}{\mbox{$\:\stackrel{<}{_{\sim}}\:$} }
\newcommand{\gsim}{\mbox{$\:\stackrel{>}{_{\sim}}\:$} }

\newcommand{\FeH}{{\rm Fe}/{\rm H}}
\newcommand{\Z}{{\rm Z}}

\begin{document}
\title{The Metallicity of Stars with Close Companions}
\author{Daniel Grether\altaffilmark{1} \& Charles H. Lineweaver\altaffilmark{2}}
\altaffiltext{1}{Department of Astrophysics, School of Physics, University of New South Wales, Sydney, NSW 2052, Australia}
\altaffiltext{2}{Planetary Science Institute, Research School of Astronomy and Astrophysics \&  
Research School of Earth Sciences, Australian National University, Canberra, ACT, Australia}


\begin{abstract}
We examine the relationship between the frequency of close companions (stellar and planetary companions with orbital 
periods $ < 5$ years) and the metallicity of their Sun-like ($\sim$ FGK) hosts. We confirm and quantify a $\sim 4 \sigma$ 
positive correlation between host metallicity and planetary companions. We find little or no dependence on spectral 
type or distance in this correlation. In contrast to the metallicity dependence of planetary companions, stellar 
companions tend to be more abundant around low metallicity hosts. At the $\sim 2 \sigma$ level we find an anti-correlation 
between host metallicity and the presence of a stellar companion. Upon dividing our sample into FG and K sub-samples, 
we find a negligible anti-correlation in the FG sub-sample and a $\sim 3 \sigma$ anti-correlation in the K sub-sample. 
A kinematic analysis suggests that this anti-correlation is produced by a combination of low-metallicity, high-binarity 
thick disk stars and higher-metallicity, lower-binarity thin disk stars.
\end{abstract}


\keywords{binaries: close -- stars: abundances -- stars: kinematics}


\section{Introduction}

With the detection to date of more than $160$ exoplanets using the Doppler technique, the observation of \citet{Gonzalez97} 
that giant close--orbiting exoplanets have host stars with relatively high stellar metallicity compared to the average 
field star has gotten stronger \citep{Reid02, Santos04, Fischer05, Bond06}. To understand the nature of this correlation 
between high host metallicity and the presence of Doppler-detectable exoplanets, we investigate whether this correlation 
extends to stellar mass companions.

There has been a widely held view that metal-poor stellar populations possess few stellar companions 
\citep{Batten73, Latham88, Latham04}. This may have been largely due to the difficulty of finding binary stars in the 
galactic halo, e.g. \citet{Gunn79}. \citet{Duquennoy91} investigated the properties of stellar companions amongst Sun-like 
stars but did not report a relationship between stellar companions and host metallicity. \citet{Latham02} and \citet{Carney05} 
reported a lower binarity for stars on retrograde Galactic orbits compared to stars on prograde Galactic orbits but found 
no dependence between binarity and metallicity within those two kinematic groups. \citet{Dall05} speculated that the 
frequency of host stars with stellar companions may be correlated with metallicity in the same way that host stars with 
planets are.

In this paper we describe and characterize the correlation between host metallicity and the fraction of planetary and 
stellar companions. In Section \ref{sec:sample} we define our sample of close planetary and stellar companions and we 
describe the variety of techniques used to obtain metallicities of stars that do not have spectroscopic metallicities 
from Doppler searches. In Section \ref{sec:analysis} we analyze the distribution of planetary and stellar companions as 
a function of host metallicity. We confirm and quantify the correlation between planet-hosts and high metallicity and 
we find a new anti-correlation between the frequency of stellar companions and high metallicity. In Section \ref{sec:binary} 
we compare our stellar companion results to analogous analyses of the \citet{Nordstrom04} and \citet{Carney05} samples.


\section{The Sample}
\label{sec:sample}

We analyze the distribution of the metallicities of FGK main-sequence stars with close companions (period $< 5$ years).
For this we use the sample of stars analyzed by \citet{Grether06}. This subset of `Sun-like' stars in the Hipparcos 
catalog, is defined by $0.5 \leq B-V \leq 1.0$ and $5.4(B-V) + 2.0 \leq M_V \leq 5.4(B-V) - 0.5$. This forms a 
parallelogram -0.5 and 2.0 mag, below and above an average main-sequence in the HR diagram. The stars range in spectral 
type from approximately F7 to K3 and in absolute magnitude in V band from 2.2 to 7.4. From this we define a more complete 
closer ($d < 25$ pc) sample of stars and an independent more distant ($25 < d < 50$ pc) sample. See \citet{Grether06} for 
additional details about the sample definition.


\subsection{Measuring Stellar Metallicity}

The metallicity of most of the extrasolar planet hosts have been determined spectroscopically. We analyse the metallicity 
data from three of these groups: (1) the McDonald observatory (hereafter, McD) group \citep[e.g.][]{Gonzalez01, Laws03}, 
(2) the European Southern Observatory (hereafter, ESO) group \citep[e.g.][]{Santos04, Santos05}, and (3) the Keck, Lick 
and Anglo-Australian observatory (hereafter, KLA) group \citep{Fischer05, Valenti05}.  

\begin{figure}[!t]
\epsscale{1.1}
\plotone{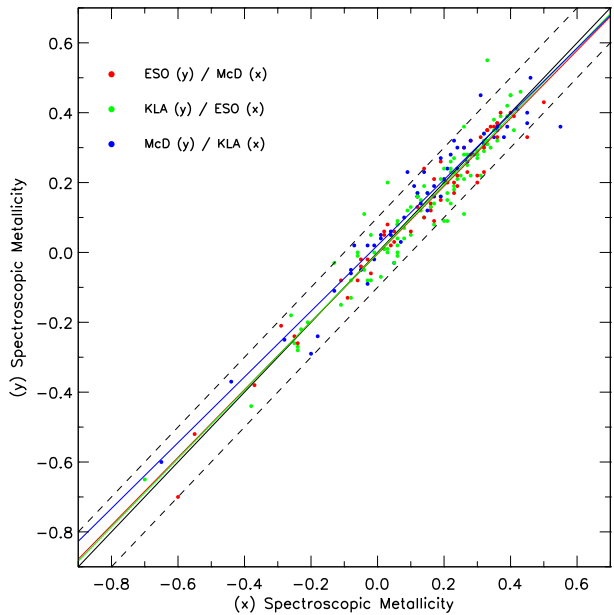}
\caption{Exoplanet Target Stars Metallicity Comparison. We compare the spectroscopic exoplanet target metallicities of the 
McD, ESO and KLA groups. The 59 red dots compare the ESO to the McD values of exoplanet target metallicity that these groups 
have in common. We find that the ESO values are on average $0.01$ dex smaller than the McD values with a dispersion of $0.05$ dex. 
Similarly the 99 green dots compare the KLA values to the average $0.01$ dex smaller ESO values with a dispersion of $0.06$ dex.
The 56 blue dots compare the KLA to the average $0.01$ dex larger McD values with a dispersion of $0.06$ dex. A solid black 
line shows the slope-one line with dashed lines at $\pm 0.1$ dex. The three linear best-fits for these three comparisons are 
nearly identical to the slope-one line and almost all scatter is contained within 0.1 dex. The relationship between the McD and 
ESO values is very close with a marginally looser relationship to the KLA values. Thus, these values for exoplanet target 
metallicity are consistent at the $\sim 0.1$ dex level.
}
\label{fig:EP_FeH_compare}
\end{figure}

\begin{figure}[!t]
\epsscale{1.1}
\plotone{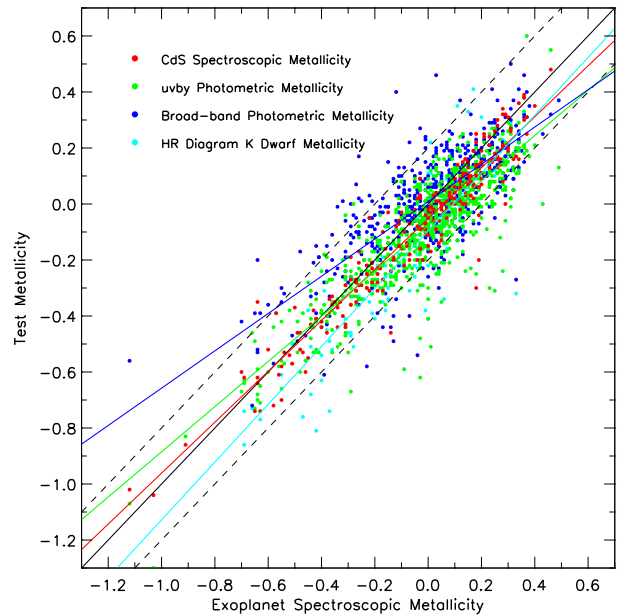}
\caption{
Metallicity values from exoplanet spectroscopy compared to four other methods of obtaining stellar metallicities. We 
compare the exoplanet target spectroscopic metallicities (plotted on the $x$ axis as a `reference') with the following 
test samples plotted on the $y$ axis: (1) CdS spectroscopic metallicities (red dots), (2) \textit{uvby} photometric 
metallicities (green dots) (3) broad-band photometric metallicities (blue dots) and (4) HR diagram K dwarf metallicities 
(aqua dots). The mean differences between the test and the reference sample metallicities ($[\FeH]_{\rm test} - [\FeH]_{\rm ref}$) 
are $-0.05$, $-0.08$, $0.01$, and $-0.10$ dex respectively, with dispersions of $0.08$, $0.11$, $0.14$ and $0.14$ dex respectively.
Comparing these mean differences and dispersions we find that the mean differences are within $1 \sigma$ of the 
solid black slope-one line and thus we regard the systematic offsets as marginal. The four linear best-fits for these 
four comparisons (shown by the four colored lines) do not show significant deviation from the slope-one line (black) 
except for the metallicities derived using broad-band photometry (dark blue line).
}
\label{fig:FeH_compare}
\end{figure}

All three of these groups find similar metallicities for the extrasolar planet target stars that they have all observed 
as shown by the comparisons in Fig. \ref{fig:EP_FeH_compare}. Apart from the $\sim 1000$ KLA target stars analyzed with a 
consistently high precision by \citet{Valenti05}, many nearby ($d < 50$ pc) FGK stars lack precise metallicities if they 
have any published measurement at all. A smaller sample of precise spectroscopic metallicities has also been published 
by the ESO group for non-planet hosting stars \citep{Santos05}.

Since the large sample of KLA stars has been taken from exoplanet target lists it also has the same biases. This includes 
selection effects (i) against high stellar chromospheric activity (ii) towards more metal-rich stars that have a greater 
probability of being a planetary host and (iii) against most stars with known close ($\theta < 2^{\prime \prime}$) stellar 
companions. We need to correct for or minimize these biases to determine quantitatively not only how the planetary distribution 
varies with host metallicity but also how the close stellar companion distribution varies with host metallicity, that is, 
we need metallicities of all stars in our sample in order to compare companion-hosting stars to non-companion-hosting stars, 
and to compare the metallicities of planet--hosting stars to the metallicities of stellar--companion--hosting stars.    

In addition to the metallicities reported by the McD, ESO and KLA groups, we use a variety of other sources and techniques 
to determine stellar metallicity although with somewhat less precision. These include other sources of spectroscopic 
metallicities such as the \citet{CdS01} (hereafter, CdS) catalog, metallicities derived from \textit{uvby} narrow-band 
photometry or broad-band photometry and metallicities derived from a star's position in the HR diagram. The precision of 
the spectroscopic metallicity values in the CdS catalog are not well quantified. However, many of the stars in the catalog 
have several independent metallicity values which we average, excluding obvious outliers. To derive metallicities from 
\textit{uvby} narrow-band photometry we apply the calibration of \citet{Martell04} to the \citet{uvby} catalog. We also use 
values of metallicity derived from broad-band photometry in \citep{Ammons06}. For stars with $5.5 < M_V < 7.3$ (K dwarfs) 
the relationship between stellar luminosity and metallicity is very tight \citep{Kotoneva02}. Using this relationship, we 
derive metallicities for some K dwarfs from their position in the HR diagram. 


\begin{deluxetable*}{lccccccl}
\tablewidth{18cm}
\tablecaption{Stellar Samples Used in Our Analysis\label{table:samples}}

\tablehead{\colhead{} & \colhead{} & \colhead{Range} & \multicolumn{5}{c}{Stars with [Fe/H] Measurements} \\
\cline{4-8}
\colhead{Sample} & \colhead{$B-V$} & \colhead{(pc)} & \colhead{Total\tablenotemark{a}} & \colhead{Binary\tablenotemark{b}} 
& \colhead{Targets\tablenotemark{c}} & \colhead{Planet Hosts\tablenotemark{d}} & \colhead{[Fe/H] Source}}

\startdata
Our FGK & $0.5-1.0$  & $d < 25$      & 453  & 45 (9.9\%)  & 379 (84\%)  & 19 (5.0\%) & Mostly Spec.\tablenotemark{e} \\
        & $0.5-1.0$  & $25 < d < 50$ & 2745 & 107 (3.9\%) & 1597 (58\%) & 36 (2.3\%) & Mostly Phot.\tablenotemark{f} \\
Our FG  & $0.5-0.75$ & $d < 25$      & 257  & 27 (10.5\%) & 228 (89\%)  & 13 (5.7\%) & Mostly Spec.\tablenotemark{g} \\
        & $0.5-0.75$ & $25 < d < 50$ & 1762 & 76 (4.3\%)  & 1167 (66\%) & 32 (2.7\%) & Mostly Phot.\tablenotemark{h} \\
Our K   & $0.75-1.0$ & $d < 25$      & 196  & 18 (9.2\%)  & 151 (77\%)  & 6 (4.0\%)  & Mostly Spec.\tablenotemark{i} \\
        & $0.75-1.0$ & $25 < d < 50$ & 983  & 31 (3.2\%)  & 430 (44\%)  & 4 (0.9\%)  & Mostly Phot.\tablenotemark{j} \\
\tableline
GC\tablenotemark{k} FGK & $0.3-1.0$  & $d < 40$ & 1375 & 378 (27.5\%) & -- & -- & \textit{uvby} Phot. \\
             & $0.5-1.0$  & $d < 40$ & 1289 & 346 (26.8\%) & -- & -- & \textit{uvby} Phot. \\
GC FG        & $0.5-0.75$ & $d < 40$ & 1117 & 291 (26.1\%) & -- & -- & \textit{uvby} Phot. \\
GC K         & $0.75-1.0$ & $d < 40$ & 172  & 55 (32.0\%)  & -- & -- & \textit{uvby} Phot. \\
\tableline
CL\tablenotemark{l} AFGK & $0.0 - 1.0$ & -- & 963 & 254 (26.4\%) & -- & -- & Spec. 
\enddata
\tablenotetext{a}{Total number of Hipparcos sun-like stars (``H" in Fig. \ref{fig:FeH_Hist}).} 
\tablenotetext{b}{Subset of total stars that are hosts to stellar companions (``S" in Fig. \ref{fig:FeH_Hist}). The percentages given 
correspond to the fraction S/H.}
\tablenotetext{c}{Subset of total stars that are exoplanet target stars (``T" in Fig. \ref{fig:FeH_Hist}). The percentages given 
correspond to the fraction T/H.}
\tablenotetext{d}{Subset of target stars that are exoplanet hosts (``P" in Fig. \ref{fig:FeH_Hist}). The percentages given 
correspond to the fraction P/T.}
\tablecomments{For notes $e$ - $j$, ``HP Spec." is high precision exoplanet target spectroscopy, ``CdS Spec." is \citet{CdS01} spectroscopy, 
``\textit{uvby} Phot." is \textit{uvby} photometry, ``BB Phot." is broad-band photometry and ``HR K Dwarf" is the method for obtaining 
metallicities for K dwarfs from their position in the HR diagram \citep{Kotoneva02}.}
\tablenotetext{e}{63\% HP Spec., 12\% CdS Spec., 20\% \textit{uvby} Phot., 1\% BB Phot. and 4\% HR K Dwarf.}
\tablenotetext{f}{19\% HP Spec., 5\% CdS Spec., 55\% \textit{uvby} Phot., 17\% BB Phot. and 4\% HR K Dwarf.}
\tablenotetext{g}{63\% HP Spec., 18\% CdS Spec., 19\% \textit{uvby} Phot.}
\tablenotetext{h}{26\% HP Spec., 7\% CdS Spec., 61\% \textit{uvby} Phot., 6\% BB Phot. and $<1\%$ HR K Dwarf.}
\tablenotetext{i}{64\% HP Spec., 5\% CdS Spec., 20\% \textit{uvby} Phot., 3\% BB Phot. and 8\% HR K Dwarf.}
\tablenotetext{j}{6\% HP Spec., 3\% CdS Spec., 44\% \textit{uvby} Phot., 36\% BB Phot. and 11\% HR K Dwarf.}
\tablenotetext{k}{``GC" is the Geneva-Copenhagen survey of the Solar neighbourhood sample \citep{Nordstrom04}. We only include those 
binaries observed by CORAVEL between 2 and 10 times (see Section \ref{sec:binary}).}
\tablenotetext{l}{``CL" is the Carney-Latham survey of proper-motion stars \citep{Carney05}. We only include those stars on prograde 
Galactic orbits ($V > -220$ km/s). The CL sample also includes 231 stars from the sample of \citet{Ryan89}.}
\end{deluxetable*}


To quantify the precision of their metallicities, we compare in Fig. \ref{fig:FeH_compare} the different methods of determining 
metallicity. We use the high precision exoplanet target spectroscopic metallicities from the McD, ESO and KLA surveys (or the 
average when a star has two or more values) as the reference sample. We compare these metallicities with metallicities of the 
following test samples: (1) CdS spectroscopic metallicities, (2) \textit{uvby} photometric metallicities, (3) broad-band 
photometric metallicities and (4) HR Diagram K dwarf metallicities. 

The result of this comparison is that the uncertainties associated with the high quality exoplanet target spectroscopic 
metallicities of McD, ESO and KLA groups are the smallest, with the CdS spectroscopic metallicities only slightly more uncertain. 
The uncertainties associated with the \textit{uvby} photometric metallicities are intermediate with broad-band photometric and 
HR diagram K dwarf metallicities being the least certain.


\subsection{Selection Effects and Completeness}
\label{sec:bias}

To minimize the scatter in the measurement of stellar metallicity while including as many stars in our samples as possible, 
we choose the metallicity source from one of the five groups based upon minimal dispersion. Thus we primarily use the 
spectroscopic exoplanet target metallicities. If no such value for metallicity is available for a star in our sample we 
use a spectroscopic value taken from the CdS catalog, followed by a \textit{uvby} photometric value, a broad-band 
value and lastly a HR diagram K dwarf value for the metallicity. We use the dispersions discussed above as estimates of 
the uncertainties of the metallicity measurements.

Almost all $(453/464 = 98\%)$ of the close ($d < 25$ pc) sample and $2745/2832 = 97\%$ of the more distant 
($25 < d < 50$ pc) sample thus have a value for metallicity. Given the different uncertainties associated with the five 
sources of metallicity, the close sample has more precise values of metallicity than the more distant sample. The 
dispersion for the close and far sample are 0.07 and 0.10 dex respectively. See Table \ref{table:samples} for details.

We also investigate the color or host mass dependence of the host metallicity distributions. Thus we split our close 
and far samples which are defined by $0.5 \leq B-V \leq 1.0 $ into 2 groups, those with $0.5 \leq B-V \leq 0.75$ which 
we call FG dwarfs and those with $0.75 < B-V \leq 1.0$ which we call K dwarfs. This split is shown in Table \ref{table:samples}. 
In this table we also show the total number of stars in the sample that have a known value of metallicity and the fraction 
that are close binaries, exoplanet target stars and exoplanet hosts.
 
In order to determine whether there is a real physical correlation between the presence of stellar or planetary companions 
and host metallicity we need to show that there are only negligible selection effects associated with the detection and 
measurement of these two quantities that could cause a spurious correlation. In Section \ref{sec:comp} we show that the 
planetary companion fraction should be complete for planetary companions with periods less than 5 years for the sample of 
target stars that are being monitored for exoplanets. This completeness helps assure minimal spurious correlation between 
the probability of detecting planetary companions and host metallicity. 


\begin{deluxetable*}{lcclll}
\tablecaption{Metallicity and Frequency of Hosts with Close Planetary and Stellar Companions\label{table:companions}}
\tablehead{\colhead{Companions} & \colhead{Range} & \colhead{Total} & \colhead{Metal-poor} & \colhead{Sun-like} & \colhead{Metal-rich}}
\startdata
Planets    & $d < 25$      & 19        & 2 (11\%)   & 5 (26\%)  & 12 (63\%) \\
Stars      & $d < 25$      & 45\tablenotemark{a}  & 25 (56\%)  & 14 (31\%) & 6 (13\%) \\
Planets    & $25 < d < 50$ & 36        & 3 (8\%)    & 9 (25\%)  & 24 (67\%) \\
Stars      & $25 < d < 50$ & 107\tablenotemark{b} & 55 (51\%)  & 36 (34\%) & 16 (15\%) 
\enddata
\tablenotetext{a}{An additional 2 hosts with unknown metallicity have stellar companions.}
\tablenotetext{b}{An additional 3 hosts with unknown metallicity have stellar companions.}
\end{deluxetable*}


The stellar companion sample is made up of two subsamples: those companions detected as part of an exoplanet survey and 
those that were not. The target list for exoplanets is biased against stellar binarity as discussed in \citet{Grether06}.
We show that there is negligible bias between the probability of detecting stellar companions and host metallicity in two ways: 
(1) by showing that our close sample of stellar companions is nearly complete and (2) by using the Geneva-Copenhagen 
survey (hereafter, GC) of the Solar neighbourhood \citep{Nordstrom04} sample of stars, containing similar types of stars 
to those found in our sample, as an independent check on our results.

The GC sample of stars which contains F0-K3 stars is expected to be complete for stars with stellar companions closer than 
$d < 40$ pc. For the close sample of stars ($d < 25$ pc), the northern hemisphere of stars with close stellar companions is 
approximately complete. The southern hemisphere of stars is also nearly complete if we include the binary stars from \citet{Jones02} 
that are likely to fall within our sample \citep{Grether06}. We then find that $\sim 10\%$ of stars have stellar companions 
with periods shorter than 5 years. If we make a small asymmetry correction (to account for the southern hemisphere not 
being as well monitored for binaries) we find that $\sim 11 \pm 3\%$ of stars have stellar companions within this period 
range \citep{Grether06}. We also compare our sample with that of the ``Carney-Latham" survey (hereafter, CL) of proper-motion 
stars \citep{Carney87, Carney94} in Section \ref{sec:binary}. The CL sample also contains $\sim 11\%$ of stars with stellar 
companions with periods shorter than 5 years \citep{Latham02}. We tabulate the properties of all these samples in 
Table \ref{table:samples}.


\begin{figure}[!t]
\epsscale{1.1}
\plotone{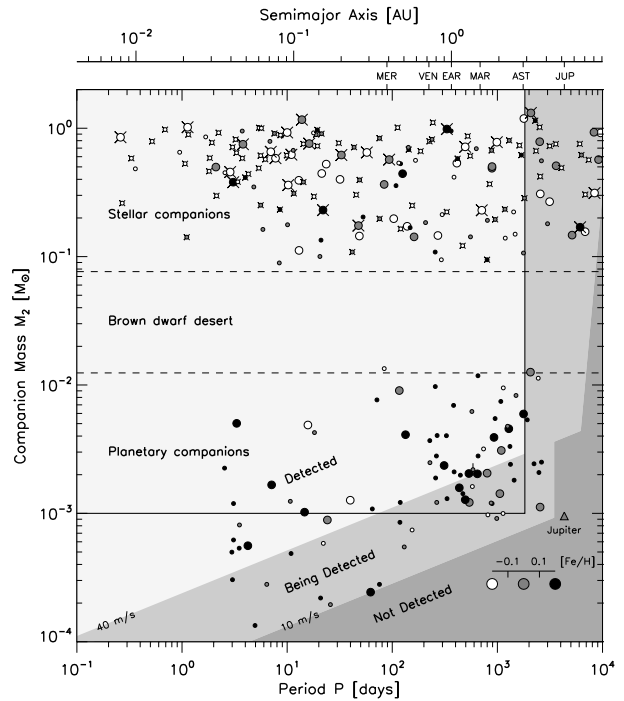}
\caption{Masses and periods of close companions to stellar hosts of FGK spectral type. We split the close companion sample 
into 3 groups defined by the metallicity of their host star: metal-poor ([\FeH] $< -0.1$), Sun-like ($-0.1 \leq$ [\FeH] $\leq 0.1$) 
and metal-rich ([\FeH] $> 0.1$) which are plotted as white, grey and black dots respectively. The larger points are companions 
orbiting stars in the more complete $d < 25$ pc sample, while the smaller points are companions to stars at distances 
between $25 < d < 50$ pc. We divide the stellar companions into those not monitored by one of the exoplanet search programs 
(shown with an `X' behind the point) and those that are monitored. Both groups of stellar companions are distributed over 
the entire less-biased region (enclosed by thick line). Hence any missing stellar companions should be randomly distributed. 
For multiple companion systems, we select the most massive companion in our less-biased sample to represent the system. 
}
\label{fig:m_P}
\end{figure}

\begin{figure}[!t]
\epsscale{1.1}
\plotone{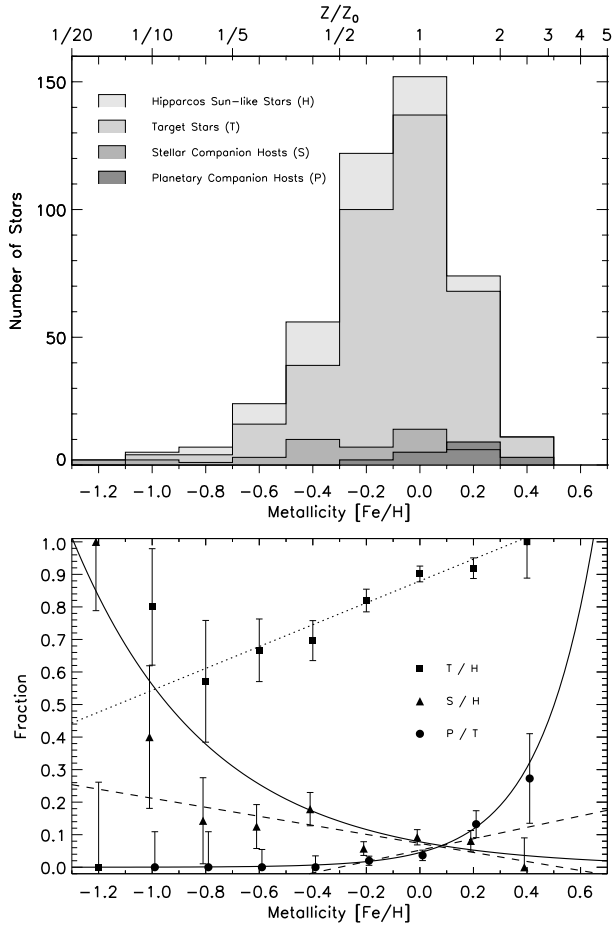}
\caption{Metallicities ($[\FeH]$) of 453 stars in our close $d < 25$ pc sample. {\it Top:} The lightest shade of grey 
are the Hipparcos `Sun-like' stars in our close sample. The next darker shade of grey are all of the stars that are 
exoplanet targets. This is followed by a still darker shade of grey which are hosts of stellar companions. The darkest 
shade of grey are exoplanet hosts. In the {\it Bottom} plot, the fraction of target stars, stellar companion hosts and 
planetary companion hosts are shown by squares, triangles and circles respectively. The linear best-fit to the target 
fraction is shown by a dotted line. The linear and exponential best-fits to the stellar and planetary companion fractions 
are shown by dashed and solid lines respectively.
}
\label{fig:FeH_Hist}
\end{figure}

\begin{figure}[!t]
\epsscale{1.1}
\plotone{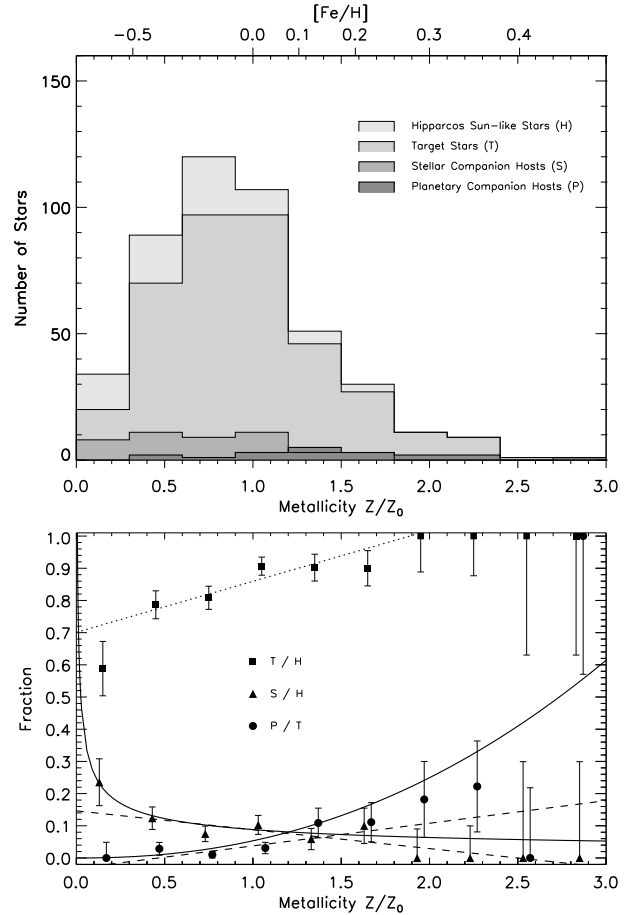}
\caption{Same as Fig. \ref{fig:FeH_Hist} except that metallicity is plotted linearly as $\Z/\Z_{\sun}$. All of the 
metal-rich ($\Z/\Z_{\sun} > 1.8)$ sample stars are being monitored for exoplanets but as the stellar metallicity 
decreases so does the fraction being monitored. This is because of a bias towards selecting more metal-rich target 
stars for observation due to an increased probability of planetary companions orbiting metal-rich host stars. The linear 
best-fit to the target fraction is shown by a dotted line. The linear and power-law best-fits to the stellar and 
planetary companion fractions are shown by dashed and solid lines respectively.
}
\label{fig:Z_Hist}
\end{figure}


\subsection{Close Companions}
\label{sec:comp}

The close companions included in our $d < 25$ pc and $25 < d < 50$ pc samples are enclosed in a rectangle of mass-period 
space shown in Fig. \ref{fig:m_P}. These companions have primarily been detected using the Doppler technique but the 
stellar companions have been detected with a variety of techniques not exclusively from high precision exoplanet Doppler 
surveys. Thus we need to consider the selection effects of the Doppler method in order to define a less-biased sample of 
companions \citep{Lineweaver03}. Given a fixed number of targets, the ``Detected" region should contain all companions 
that will be found for this region of mass-period space. The ``Being Detected" region should contain some but not all 
companions that will be found in this region and the ``Not Detected" region contains no companions since the current 
Doppler surveys are either not sensitive enough or have not been observing for a long enough duration to detect 
companions in this regime. Thus as a consequence of the exoplanet surveys' limited monitoring duration and sensitivity 
for our sample we only select those companions with an orbital period $P < 5$ years and mass $M_2 > 0.001 M_\sun$. 

In \citet{Grether06} we found that companions with a minimum mass in the brown dwarf mass regime were likely to be low mass 
stellar companions seen face on, thus producing a very dry brown dwarf desert. We also included the 14 stellar companions 
from \citet{Jones02} that have no published orbital solutions but are assumed to orbit within periods of 5 years. We find one 
new planet and no new stars in our less biased rectangle when compared with the data used in \citet{Grether06}. This new 
planet HD 20782 (HIP 15527) (indicated by a vertical line through the point in Fig. \ref{fig:m_P}), has been monitored for 
well over 5 years but only has a period of $\sim 1.6$ years and a minimum mass of $1.8 M_{\rm Jup}$ placing it just between 
the ``Detected" and ``Being Detected" regions. While most planets are detected within a time frame comparable to the 
period, the time needed to detect this planet was much longer than its period because of its unusually high eccentricity 
of 0.92 \citep{Jones06}. We thus have two groups of close companions to analyse as a function of host metallicity - giant 
planets and stars.

In Fig. \ref{fig:m_P} we split the close companion sample into 3 groups defined by the metallicity of their host star: 
metal-poor ($[\FeH] < -0.1$), Sun-like ($-0.1 \leq [\FeH] \leq 0.1$) and metal-rich ($[\FeH] > 0.1$) which are plotted as 
white, grey and black dots respectively. Fig. \ref{fig:m_P} suggests that the hosts of planetary companions are generally 
metal-rich whereas the hosts of stellar companions are generally metal-poor. Table \ref{table:companions} and 
Fig. \ref{fig:FeH_Hist} confirm the correlation between exoplanets and high-metallicity and indicate an anti-correlation 
between stellar companions and high metallicity.


\section{Close Companion - Host Metallicity Correlation}
\label{sec:analysis}

\begin{figure}[!t]
\epsscale{1.1}
\plotone{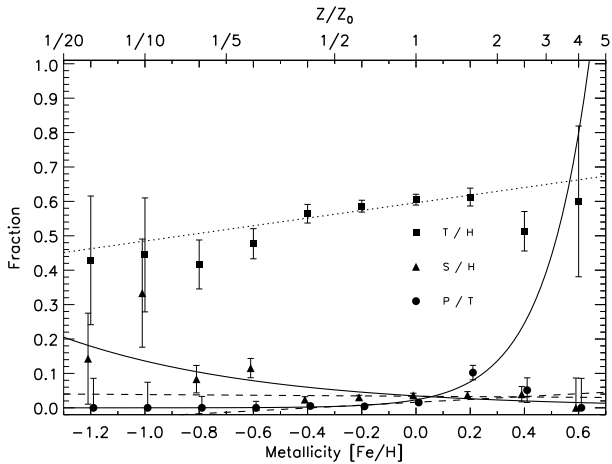}
\caption{Same as Fig. \ref{fig:FeH_Hist} except for the 2745 stars in the more distant $25 < d < 50$ pc sample. It 
is harder to detect distant planets because of signal to noise considerations which limit observations to the brighter 
stars. This fainter, more distant sample relies more on photometric metallicity determinations than does the closer, 
brighter sample which has predominantly spectroscopic metallicity determinations (see Table \ref{table:samples}). 
The fraction of stars being monitored for exoplanets is much lower than in Fig. \ref{fig:FeH_Hist}.
}
\label{fig:FeH_Hist_50}
\end{figure}

\begin{figure}[!t]
\epsscale{1.1}
\plotone{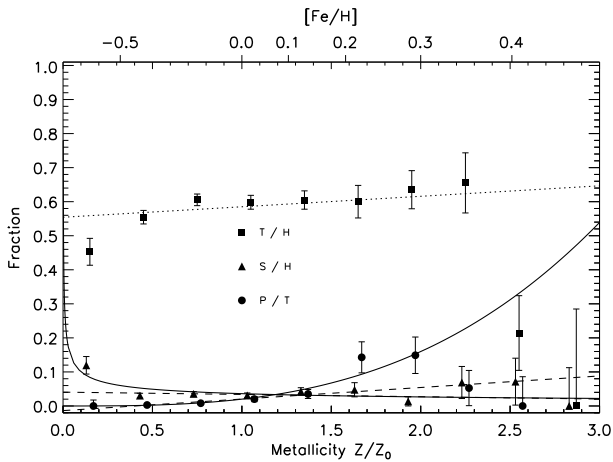}
\caption{Same as Fig. \ref{fig:FeH_Hist_50} except that metallicity is plotted linearly as $\Z/\Z_{\sun}$ analogous 
to Fig. \ref{fig:Z_Hist}. In this more distant sample we find the same trends as in Fig. \ref{fig:Z_Hist} but they 
are not as prominent.
}
\label{fig:Z_Hist_50}
\end{figure}

We examine the distribution of close companions as a function of stellar host metallicity in our two samples. We do this 
quantitatively by fitting power-law and exponential best-fits to the metallicity data expressed both linearly and 
logarithmically. We define the logarithmic $[\FeH]$ and linear $\Z/\Z_{\sun}$ metallicity as follows:   

\begin{equation}
[\FeH] = \log(\FeH) - \log(\FeH)_{\sun} = \log(\Z/\Z_{\sun})
\end{equation}

\noindent
where Fe and H are the number of iron and hydrogen atoms respectively and $\Z=\FeH$. We examine the close planetary 
companion probability $P_{\rm planet}$ and the close stellar companion probability $P_{\rm star}$ as a function of $[\FeH]$ 
in Figs. \ref{fig:FeH_Hist} and \ref{fig:FeH_Hist_50} for the $d < 25$ and $25 < d < 50$ pc samples respectively. 
Similarly we also examine $P_{\rm planet}$ and $P_{\rm star}$ as a function of $\Z/\Z_{\sun}$ in Fig. \ref{fig:Z_Hist} 
and \ref{fig:Z_Hist_50}, which is effectively just a re-binning of the data in Fig. \ref{fig:FeH_Hist} and Fig. \ref{fig:FeH_Hist_50}.
We then find the linear best-fits to the planetary and stellar companion fraction distributions as shown by the dashed lines 
in Figs. \ref{fig:FeH_Hist}-\ref{fig:Z_Hist_50}.

We also fit an exponential to the $[\FeH]$ planetary \citep[as in][]{Fischer05} and stellar companion fraction distributions 
in Figs. \ref{fig:FeH_Hist} and \ref{fig:FeH_Hist_50} and equivalently a power-law to the data points for the $\Z/\Z_{\sun}$ 
plots, Figs. \ref{fig:Z_Hist} and \ref{fig:Z_Hist_50}. The two linear parameterizations that we fit to the data are:

\begin{eqnarray}
\label{eq:linear}
P_{\rm lin\;Fe/H}    &=& a [\FeH] + P_{\sun} \\
\label{eq:linear_z}
P_{\rm lin\;Z/Z_{0}} &=& A (\Z/\Z_{\sun}) + (P_{\sun} - A)
\end{eqnarray}

\noindent
and the two non-linear parameterizations are:

\begin{eqnarray}
\label{eq:alpha}
P_{\rm EP} &=& P_{\sun} 10^{ \alpha \;[\FeH]} \\
\label{eq:alpha_z}
           &=& P_{\sun} (\Z/\Z_{\sun})^{\alpha} 
\end{eqnarray}

\noindent
where $P_{\sun}$ is the fraction of stars of solar metallicity (i.e. $[\FeH] = 0$ and $\Z/\Z_{\sun} =1$) with companions.  

If the fits for the parameters $a,A$ and $\alpha$ are consistent with zero then there is no correlation between the fraction 
of stars with companions and metallicity. On the other hand, a non-zero value, several sigma away from zero suggests a 
significant correlation ($a,A$ or $\alpha > 0$) or anti-correlation ($a,A$ or $\alpha < 0$).


The best-fit parameters $a, A$ and $P_{\sun}$ (but not $\alpha$) depend upon the period range and completeness of 
the sample. In order to compare the slopes from different samples, we parametrize this dependence 
in terms of the average companion fraction $P_{\rm avg}$ for the sample, i.e., if the average  companion fraction for a 
sample is twice as large as for another sample, the best-fit slopes $a$ and $A$ as well as the fraction of stars of solar 
metallicity $P_{\sun}$  will also be twice as large. To compare samples with different $P_{\rm avg}$, we scale the best-fit 
Eqs. \ref{eq:linear}-\ref{eq:alpha_z} to a common average companion fraction by dividing each equation by $P_{\rm avg}$. 
Thus, we scale the best-fit parameters $a, A$ and $P_{\sun}$ by dividing each by $P_{\rm avg}$. These scaled parameters 
are then referred to as $a' = a/P_{\rm avg}$, $A' = A/P_{\rm avg}$ and $P'_{\sun} = P_{\sun}/P_{\rm avg}$. We list the unscaled 
best-fit parameters $a, A, P_{\sun}, \alpha$ along with $P_{\rm avg}$ for each sample in Table \ref{table:bestfit}. The 
parameters $a'=a/P_{\rm avg}$ and $\alpha$ of the different samples are compared in Fig. \ref{fig:slopes}.

We consistently find in Figs. \ref{fig:FeH_Hist} and \ref{fig:FeH_Hist_50} that metal-rich stars are being monitored more 
extensively for exoplanets than metal-poor stars as quantified by the ``Target Stars" / ``Hipparcos Sun-like Stars" ratio. 
This is because of a bias towards selecting more metal-rich stars for observation due to an increased probability of 
planetary companions orbiting metal-rich host stars. Note that this bias is well-represented by a linear trend as shown 
by the dotted best-fit line in these figures, and is not just a case of a few high metallicity stars being added to the highest 
metallicity bins. We correct for this bias by calculating ``P/T" not ``P/H" for each metallicity bin.

We find a correlation between $[\FeH]$ and the presence of planetary companions in Fig. \ref{fig:FeH_Hist}. The 
linear best-fit (Eq. \ref{eq:linear}) has a gradient of $a = 0.18 \pm 0.07$ ($\chi^2_{\rm red} = 1.21$) and thus the correlation 
is significant at the $2 \sigma$ level. The non-linear best-fit (Eq. \ref{eq:alpha}) is $\alpha = 2.09 \pm 0.54$ 
($\chi^2_{\rm red} = 0.16$) and thus the correlation is significant at slightly more than the $3 \sigma$ level.

Similarly we find a correlation between linear metallicity $\Z/\Z_{\sun}$ and the presence of planetary companions in 
the same data re-binned in Fig. \ref{fig:Z_Hist}. The linear best-fit (Eq. \ref{eq:linear_z}) has a gradient of 
$A = 0.07 \pm 0.03$ ($\chi^2_{\rm red} = 1.25$) and the non-linear best-fit (Eq. \ref{eq:alpha_z}) has an exponent of 
$\alpha = 2.22 \pm 0.39$ ($\chi^2_{\rm red} = 1.00$) which are non-zero at the $\sim 2 \sigma$ and $\sim 5 \sigma$ significance levels 
respectively. These results are summarized in Table \ref{table:bestfit}. We can compare the non-linear best-fit (Eq. \ref{eq:alpha_z}) 
for linear metallicity $\Z/\Z_{\sun}$ and the non-linear best-fit (Eq. \ref{eq:alpha}) for log metallicity $[\FeH]$ since both 
contain the parameter $\alpha$. As shown by the $\chi^2$ per degree of freedom $\chi^2_{\rm red}$, the non-linear goodness of fit 
is better than the linear goodness of fit. We rely on the best fitting functional form which is the non-linear parameterization 
of our results although we use both parameterizations in our analysis. 

We combine these two non-independent, non-linear best-fit estimates by computing their weighted average. We assign an error 
to this average by adding in quadrature (1) the difference between the two estimates and (2) the nominal error on the average. 
Thus our best estimate is $\alpha = 2.2 \pm 0.5$. Hence the correlation between the presence of planetary companions and host 
metallicity is significant at the $\sim 4 \sigma$ level for a non-linear best-fit and at the $\sim 2 \sigma$ level with a 
lower goodness-of-fit for a linear best-fit in our close, most complete sample.

In Fig. \ref{fig:FeH_Hist}, and its re-binned equivalent Fig. \ref{fig:Z_Hist}, we find an anti-correlation between the presence 
of stellar companions and host metallicity. The linear stellar companion best-fits have gradients of $a = -0.14 \pm 0.06$ 
($\chi^2_{\rm red} = 3.00$) and $A = -0.06 \pm 0.03$ ($\chi^2_{\rm red} = 0.91$) respectively, both significant at the 
$\sim 2 \sigma$ level. The non-linear best-fit to the stellar companions as a function of $[Fe/H]$ in Fig. \ref{fig:FeH_Hist} 
is $\alpha = -0.86 \pm 0.10$ ($\chi^2_{\rm red} = 1.33$) and the non-linear best-fit to the stellar companions as a function 
of $\Z/\Z_{\sun}$ in Fig. \ref{fig:Z_Hist} is $\alpha = -0.47 \pm 0.18$ ($\chi^2_{\rm red} = 0.40$). Averaging these two as 
above we obtain, $-0.8 \pm 0.4$, which is significant at the $\sim 2 \sigma$ level. All these best-fits are summarized in 
Table \ref{table:bestfit}.


\begin{figure}[!t]
\epsscale{1.1}
\plotone{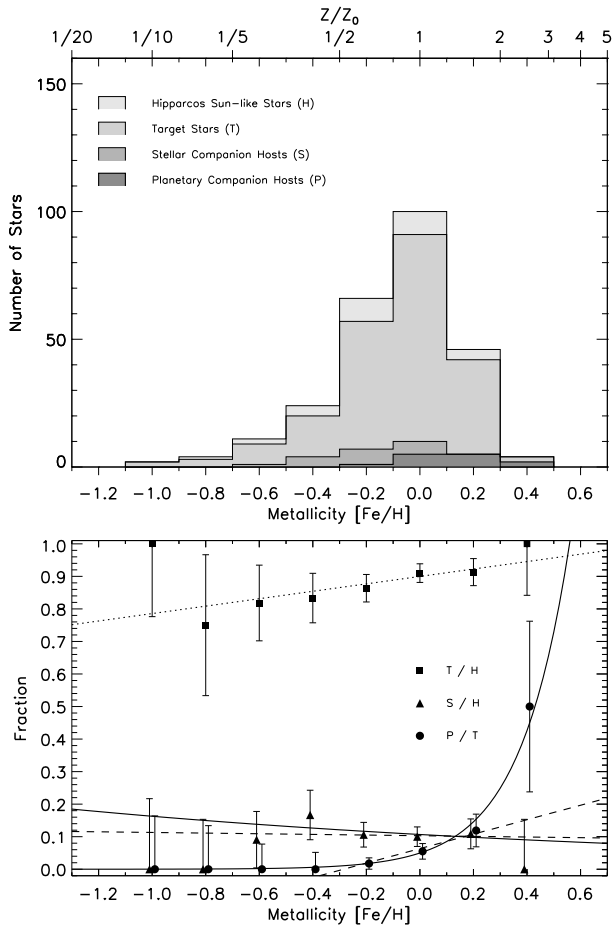}
\caption{Same as Fig. \ref{fig:FeH_Hist} for the stars in our close $d < 25$ pc sample but only for FG dwarfs 
($B-V \leq 0.75$). All stars have known metallicity in this sample. There is no apparent anti-correlation between 
metallicity and the presence of stellar companions.
}
\label{fig:FeH_Hist_FG}
\end{figure}

\begin{figure}[!t]
\epsscale{1.1}
\plotone{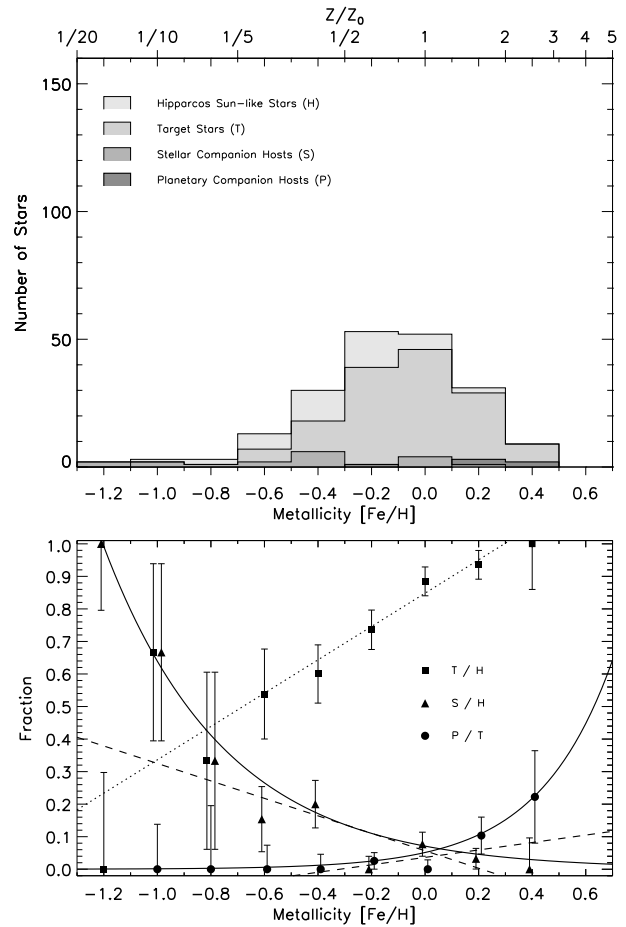}
\caption{Same as Fig. \ref{fig:FeH_Hist} for the stars in our close $d < 25$ pc sample but only for K dwarfs 
($B-V > 0.75$). This plot shows a strong anti-correlation between metallicity and the presence of stellar companions.
}
\label{fig:FeH_Hist_K}
\end{figure}


Having found a correlation for planetary companions and an anti-correlation for stellar companions in our close sample and having 
found them to be robust to different metallicity binnings, we perform various other checks to confirm their reality. We check 
the robustness of both results to (i) distance and (ii) spectral type ($\sim$ mass) of the host star.

To check if these anti-correlations have a distance dependence, we repeat this analysis for the less complete $25 < d < 50$ pc 
sample. As shown by the best-fits in Figs. \ref{fig:FeH_Hist_50} and \ref{fig:Z_Hist_50} and summarized in Table \ref{table:bestfit} 
we find only a marginal anti-correlation between the presence of stellar companions and host metallicity for the linear best-fits. 
The non-linear best-fits however still suggest an anti-correlation with $\alpha = -0.59 \pm 0.12$ for log metallicity $[\FeH]$ and 
$\alpha = -0.44 \pm 0.12$ for linear metallicity $\Z/\Z_{\sun}$, which are significant at the $4 \sigma$ and $3 \sigma$ levels respectively. 
Combining these two estimates as described above we find $\alpha = -0.5 \pm 0.2$ significant at the $2 \sigma$ level in 
the $25 < d < 50$ pc sample.

The correlation between the presence of planetary companions and host metallicity for the less complete $25 < d < 50$ pc is 
significant at the $4 \sigma$ and $3 \sigma$ levels for the linear best-fits $a$ and $A$ respectively. The non-linear best-fit 
correlation has $\alpha = 2.56 \pm 0.45$ for log metallicity $[\FeH]$ and $\alpha = 3.00 \pm 0.46$ for linear metallicity 
$\Z/\Z_{\sun}$ which are significant at the $5 \sigma$ and $6 \sigma$ levels respectively. Combining these two estimates we find 
the weighted average as above of $\alpha = 2.8 \pm 0.6$ significant at the $4 \sigma$ level.

Having found the correlation for planetary companions and the anti-correlation for stellar companions robust to binning but less 
robust in the less-complete more distant sample, we test for spectral type ($\sim$ host mass) dependence. We split our sample into 
bluer and redder subsamples to investigate the effect of spectral type on the close-companion/host-metallicity relationship. 
We define the bluer subsample by $B-V \leq 0.75$ (FG spectral type stars) and the redder subsample by $B-V > 0.75$ (K spectral 
type stars). Since $B-V$ has a metallicity dependence, a cut in $B-V$ will not be a true mass cut, but a diagonal cut in 
mass vs metallicity. Thus, interpreting a $B-V$ cut as a pure cut in mass introduces a spurious anti-correlation between mass 
and metallicity.


\begin{deluxetable*}{lccccccccl}
\tablewidth{18cm}
\tablecaption{Best-fit Trends for Close--Companion Host--Metallicity Correlation (see Fig. \ref{fig:slopes})\label{table:bestfit}}

\tablehead{\colhead{} & \colhead{Range} & \colhead{} & \colhead{} & \colhead{} & \multicolumn{2}{c}{Linear} & 
\multicolumn{2}{c}{Non-Linear} & \colhead{} \\
\cline{6-7} \cline{8-9}
\colhead{Sample} & \colhead{(pc)} & \colhead{Type\tablenotemark{a}} & \colhead{Figure} & \colhead{Companions} & 
\colhead{$a$ or $A$} & \colhead{$P_{\sun}$ [\%]} & \colhead{$\alpha$} & \colhead{$P_{\sun}$ [\%]} & \colhead{${P_{\rm avg}}$\tablenotemark{b} [\%]}}

\startdata
    & $d < 25$ & $[\FeH]$ & Fig. \ref{fig:FeH_Hist} & Planets & $0.18 \pm 0.07$  & $4.8 \pm 1.2$ & $2.09 \pm 0.54$  & $4.5 \pm 1.3$ & $5.0$ \\
    & $d < 25$ & $[\FeH]$ & Fig. \ref{fig:FeH_Hist} & Stars   & $-0.14 \pm 0.06$ & $7.3 \pm 1.4$ & $-0.86 \pm 0.10$ & $7.8 \pm 1.3$ & $9.9$ \\
    & $d < 25$ & $\Z/\Z_{\sun}$ & Fig. \ref{fig:Z_Hist} & Planets & $0.07 \pm 0.03$  & $3.8 \pm 2.5$ & $2.22 \pm 0.39$  & $5.3 \pm 1.4$ & $5.0$ \\
    & $d < 25$ & $\Z/\Z_{\sun}$ & Fig. \ref{fig:Z_Hist} & Stars   & $-0.06 \pm 0.03$ & $8.8 \pm 3.5$ & $-0.47 \pm 0.18$ & $8.8 \pm 1.5$ & $9.9$ \\
    & $d < 25$ & Avg. & -- & Planets & -- & -- & $2.2 \pm 0.5$  & -- & -- \\
Our & $d < 25$ & Avg. & -- & Stars   & -- & -- & $-0.8 \pm 0.4$ & -- & -- \\
FGK & $25 < d < 50$ & $[Fe/H]$ & Fig. \ref{fig:FeH_Hist_50} & Planets & $0.04 \pm 0.01$  & $1.6 \pm 0.4$ & $2.56 \pm 0.45$  & $2.3 \pm 0.5$ & $2.3$ \\
    & $25 < d < 50$ & $[Fe/H]$ & Fig. \ref{fig:FeH_Hist_50} & Stars   & $-0.00 \pm 0.02$ & $3.3 \pm 0.4$ & $-0.59 \pm 0.12$ & $3.5 \pm 0.4$ & $3.9$ \\
    & $25 < d < 50$ & $Z/Z_{\sun}$ & Fig. \ref{fig:Z_Hist_50} & Planets & $0.03 \pm 0.01$  & $2.0 \pm 0.6$ & $3.00 \pm 0.46$  & $2.0 \pm 0.4$ & $2.3$ \\
    & $25 < d < 50$ & $Z/Z_{\sun}$ & Fig. \ref{fig:Z_Hist_50} & Stars   & $-0.01 \pm 0.01$ & $3.4 \pm 0.8$ & $-0.44 \pm 0.12$ & $3.6 \pm 0.4$ & $3.9$ \\
    & $25 < d < 50$ & Avg. & -- & Planets & -- & -- & $2.8 \pm 0.6$  & -- & -- \\
    & $25 < d < 50$ & Avg. & -- & Stars   & -- & -- & $-0.5 \pm 0.2$ & -- & -- \\
\tableline
       & $d < 25$      & $[Fe/H]$ & Fig. \ref{fig:FeH_Hist_FG} & Planets & $0.22 \pm 0.09$  & $6.3 \pm 1.7$  & $2.33 \pm 0.62$  & $5.0 \pm 1.7$  & $5.7$ \\
Our    & $d < 25$      & $[Fe/H]$ & Fig. \ref{fig:FeH_Hist_FG} & Stars   & $-0.01 \pm 0.08$ & $10.3 \pm 2.0$ & $-0.18 \pm 0.37$ & $10.7 \pm 2.1$ & $10.5$ \\
FG     & $25 < d < 50$ & $[Fe/H]$ & --                         & Planets & $0.05 \pm 0.02$  & $2.0 \pm 0.5$ & $2.63 \pm 0.46$  & $2.8 \pm 0.6$ & $2.7$ \\
       & $25 < d < 50$ & $[Fe/H]$ & --                         & Stars   & $0.04 \pm 0.02$  & $4.4 \pm 0.6$ & $-0.09 \pm 0.18$ & $4.7 \pm 0.6$ & $4.3$ \\
\tableline
      & $d < 25$      & $[Fe/H]$ & Fig. \ref{fig:FeH_Hist_K} & Planets & $0.11 \pm 0.10$  & $3.6 \pm 2.0$ & $1.57 \pm 0.85$  & $5.1 \pm 3.1$ & $4.0$ \\
Our   & $d < 25$      & $[Fe/H]$ & Fig. \ref{fig:FeH_Hist_K} & Stars   & $-0.27 \pm 0.07$ & $5.7 \pm 1.9$ & $-0.95 \pm 0.14$ & $7.1 \pm 2.3$ & $9.2$ \\
K     & $25 < d < 50$ & $[Fe/H]$ & --                        & Planets & $0.03 \pm 0.03$  & $0.9 \pm 0.6$ & $2.10 \pm 2.16$  & $1.3 \pm 1.0$ & $0.9$ \\
      & $25 < d < 50$ & $[Fe/H]$ & --                        & Stars   & $-0.06 \pm 0.02$ & $2.0 \pm 0.5$ & $-1.12 \pm 0.19$ & $2.0 \pm 0.5$ & $3.2$ \\
\tableline
GC\tablenotemark{c} FGK & $d < 40$ & $[Fe/H]$ & Fig. \ref{fig:FeH_N04} & Stars & $-0.15 \pm 0.05$ & $24.0 \pm 1.5$ & $-0.28 \pm 0.07$ & $23.9 \pm 1.4$ & $26.8$ \\
\tableline
GC FG & $d < 40$ & $[Fe/H]$ & Fig. \ref{fig:FeH_N04_FG} & Stars & $-0.05 \pm 0.06$ & $25.1 \pm 1.6$ & $-0.08 \pm 0.10$ & $25.5 \pm 1.6$ & $26.1$ \\
\tableline
GC K & $d < 40$ & $[Fe/H]$ & Fig. \ref{fig:FeH_N04_K} & Stars & $-0.46 \pm 0.10$ & $22.3 \pm 3.6$ & $-0.52 \pm 0.11$ & $23.7 \pm 3.9$ & $32.0$ \\
\tableline
CL\tablenotemark{d} AFGK & -- & $[Fe/H]$ & -- & Stars & $-0.07 \pm 0.06$ & $21.3 \pm 3.4$ & $-0.12 \pm 0.09$ & $21.6 \pm 3.1$ & $25.7$ \\
\tableline
Combined\tablenotemark{e} & -- & $[Fe/H]$ & Fig. \ref{fig:Carney_FeH} & Stars & $-0.10 \pm 0.03$ & $20.5 \pm 1.3$ & $-0.22 \pm 0.05$ & $20.7 \pm 1.2$ & $25.7$ 
\enddata

\tablenotetext{a}{``Type" refers to whether the data is binned in $[\FeH]$ or in $\Z/\Z_{\sun}$. For data binned in $[\FeH]$ the linear 
slope is $a$ and for data binned in $\Z/\Z_{\sun}$ the linear slope is $A$ (see Eqs. \ref{eq:linear}-\ref{eq:alpha_z}).}
\tablenotetext{b}{``$P_{\rm avg}$" is defined as the number of stars with stellar companions divided by the total number of stars. We 
use this parameter to scale $a, A$ and $P_{\sun}$ which are then referred to by $a' = a/P_{\rm avg}$, $A' = A/P_{\rm avg}$ 
and $P'_{\sun} = P_{\sun}/P_{\rm avg}$ and can be compared between different samples (see text Section \ref{sec:analysis}).}
\tablenotetext{c}{``GC" is the Geneva-Copenhagen survey of the Solar neighbourhood sample \citep{Nordstrom04}. We only include those 
binaries observed by CORAVEL between 2 and 10 times (see Section \ref{sec:binary}).}
\tablenotetext{d}{``CL" is the Carney-Latham survey of proper-motion stars \citep{Carney05}. We only include those stars on prograde 
Galactic orbits ($V > -220$ km/s) and with $[\FeH] > -1.3$. This also includes stars from the sample of \citet{Ryan89}.}
\tablenotetext{e}{Combined sample of (i) our ($d < 25$ pc F7-K3, Fig. \ref{fig:FeH_Hist}) sample, (ii) the volume-limited GC 
($d < 40$ pc F7-K3, Fig. \ref{fig:FeH_N04}) sample and (iii) the prograde Galactic orbits from \citet{Carney05}.  
Fig. \ref{fig:Carney_FeH} shows all three datasets.}
\end{deluxetable*}


The linear best-fit to the stellar companions of the FG sample ($d < 25$ pc) has a normalised gradient of 
$a' = (-0.01 \pm 0.08)/10.5\% = -0.1 \pm 0.8$ and the non-linear best-fit is $\alpha = -0.2 \pm 0.4$ as shown in 
Fig. \ref{fig:FeH_Hist_FG}. Both of these best-fits are consistent with the frequency of stellar companions being independent 
of host metallicity. The linear best-fit to the stellar companions of the K sample has a gradient of $a' = (-0.27 \pm 0.07)/9.2\% = 2.9 \pm 0.8$ 
and the non-linear best-fit is $\alpha = -1.0 \pm 0.1$ as shown in Fig. \ref{fig:FeH_Hist_K} for the close $d < 25$ pc stars. 
Both of these best-fits show an anti-correlation between the presence of stellar companions and host metallicity at above the 
$3 \sigma$ level. Less significant results are obtained for the $25 < d < 50$ FG and K spectral type samples with stellar 
companions as shown in Table \ref{table:bestfit}. 

These results suggest that the observed anti-correlation between close binarity and host metallicity is either (i) real 
and stronger for K spectral type stars than for FG stars or (ii) due to a spectral-type dependent selection effect.

Under the hypothesis that the anti-correlation between host metallicity and binarity is real for K dwarfs, there is a possible 
selection effect limited to F and G stars that could explain why we do not see the anti-correlation as strongly in them. Doppler 
broadening of the line profile, due to random thermal motion in the stellar atmosphere and stellar rotation, both increase in 
more massive F and G stars due to their higher effective temperature and faster rotation speeds compared with less massive K 
stars. This wider line profile for F and G stars results in fewer observable shifting lines thus lowering the spectroscopic binary 
detection efficiency. However we directly examine the stellar companion fraction as a function of spectral type or color $B-V$ 
in Fig. \ref{fig:BV}. For both single-lined and double-lined spectroscopic binaries, if the binary detection efficiency was 
systemically higher for K dwarfs then the anti-correlation could be a selection effect. However we find that it is fairly 
independent of spectral type. Thus the anti-correlation does not appear to be a spectral-type dependent selection effect.


\begin{figure*}[!t]
\epsscale{1.1}
\plotone{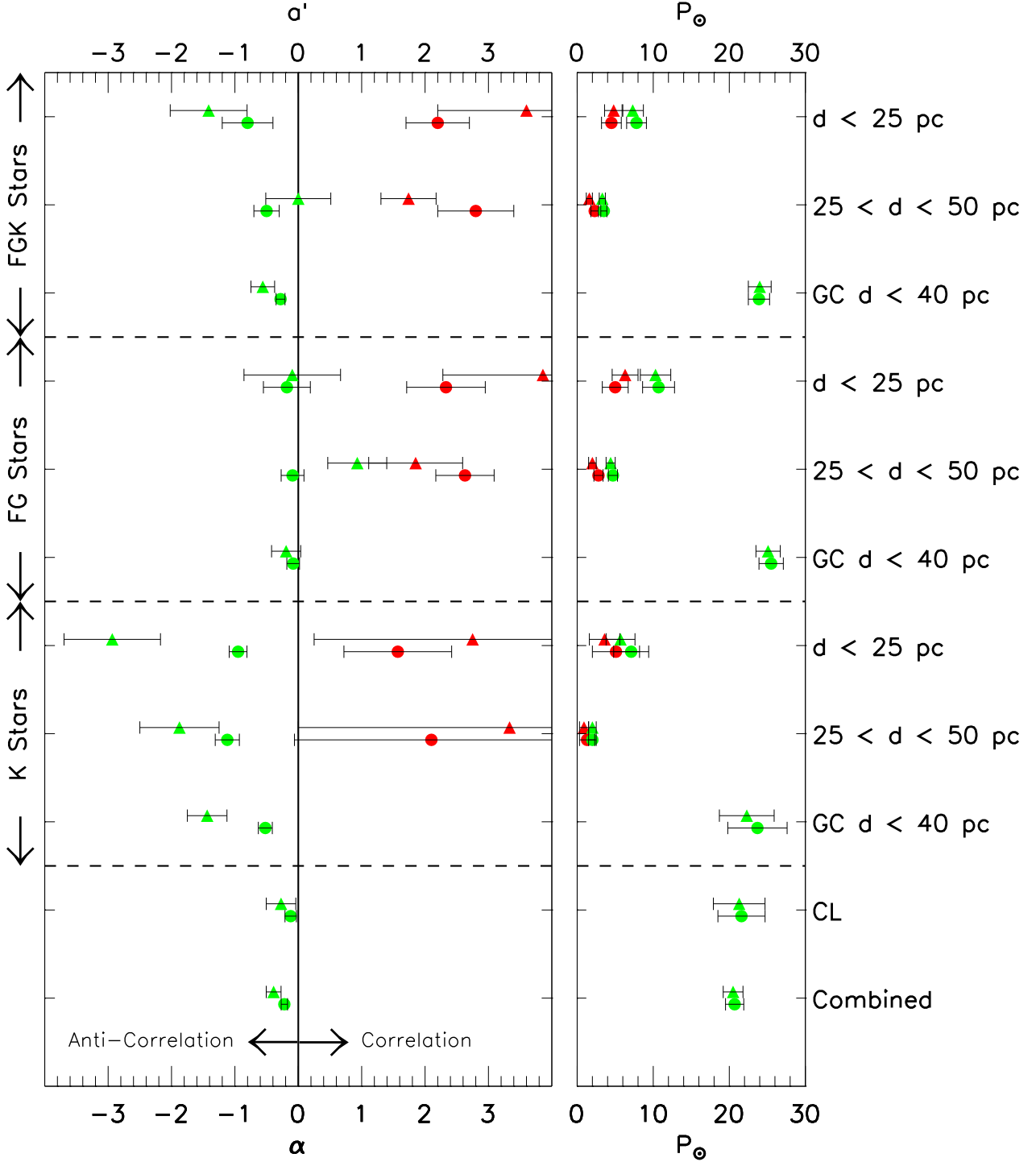}
\caption{We compare the linear $a'$ (triangles) and non-linear $\alpha$ (circles) parameterizations for the various samples 
listed in Table \ref{table:bestfit}. The red points are the best-fits to planetary companions and the green points the 
best-fits to stellar companions. The fact that the red planet values for $\alpha$ are significantly larger than zero 
confirms and quantifies the metallicity/planet correlation. The fact that the green stellar values for $\alpha$ are 
predominantly less than zero, significantly so only for K dwarfs, is a surprising new result. The labels on the RHS refer to 
the samples for which the best-fit parameterizations are valid. We normalize the linear parametrization by dividing the 
best-fit gradient $a$ by the average companion fraction $P_{\rm avg}$ (see text). This plot is a graphical version of 
Table \ref{table:bestfit} whose notes also apply to this plot. All of the best-fits are from $[\FeH]$ plots except for our 
FGK stars where $\alpha$ is the average of the best-fits to both the $[\FeH]$ and $\Z/\Z_{\sun}$ plots. The $P_{\sun}$ 
values plotted in the vertical panel on the right refer to the corresponding best-fit normalization at solar metallicity 
(Eqs. \ref{eq:linear}-\ref{eq:alpha_z}).
}
\label{fig:slopes}
\end{figure*}

\begin{figure}[!t]
\epsscale{1.1}
\plotone{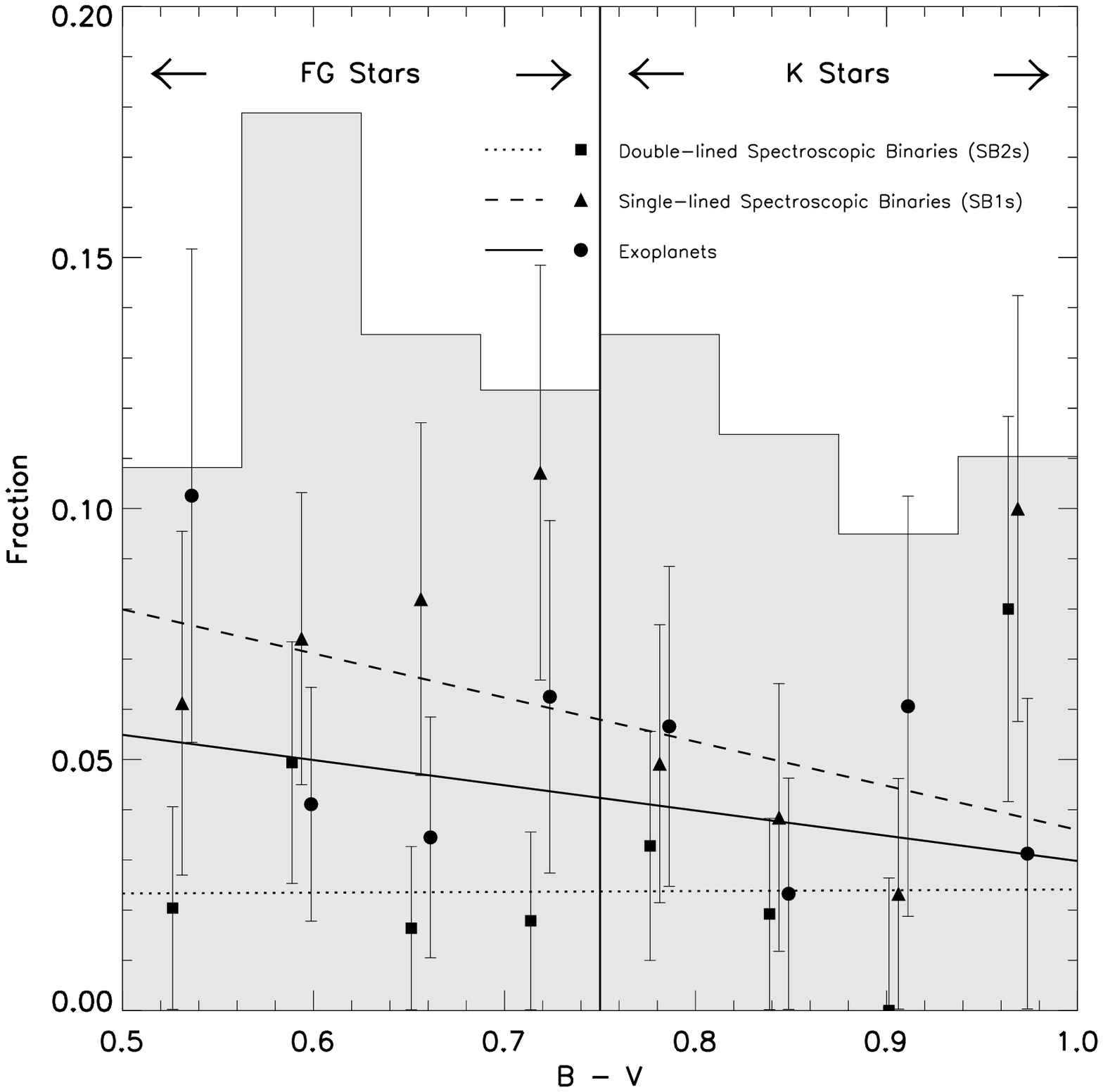}
\caption{Color ($B-V$) distribution for double-lined (squares) and single-lined (triangles) spectroscopic binaries 
(SB2s and SB1s respectively) and exoplanets (circles) in our close $d < 25$ pc sample. The linear best-fit gradient for 
SB2s is $0.00 \pm 0.06$, for SB1s it is $-0.08 \pm 0.08$ and for exoplanets it is $-0.05 \pm 0.08$. All three of these 
gradients are only significant at $\lsim 1 \sigma$ level. There is no significant correlation between SB1, SB2 or 
planetary fraction for either FG ($B-V \leq 0.75$) stars or for K ($B-V > 0.75$) stars. 
}
\label{fig:BV}
\end{figure}


We also examine the spectral type ($\sim$ mass) dependence of the correlation between planetary companions and host metallicity.
The linear best-fit to the planetary companions of the FG sample has a gradient of $a' = (0.22 \pm 0.09)/5.7\% = 3.9 \pm 1.6$ and 
the non-linear best-fit is $\alpha = 2.3 \pm 0.6$ as shown in Fig. \ref{fig:FeH_Hist_FG} for the close $d < 25$ pc stars. 
These are significant at the 2 and 3 $\sigma$ levels respectively. The linear best-fit to the planetary companions of the 
K sample has a gradient of $a' = (0.11 \pm 0.10)/4.0\% = 2.8 \pm 2.5$ and the non-linear best-fit is $\alpha = 1.6 \pm 0.9$ 
as shown in Fig. \ref{fig:FeH_Hist_K} for the close $d < 25$ pc stars. These are both significant at between the 1 and 
2 $\sigma$ levels. The K sample contains fewer planetary and stellar companions compared to the FG sample. Both the linear 
and non-linear fits are consistent between the FG and K samples suggesting that the correlation between the presence of 
planetary companions and host metallicity is independent of spectral type and consequently host mass. The fraction of 
planetary companions is also fairly independent of spectral type as shown in Fig. \ref{fig:BV}. 

Thus our results suggest that the correlation between the presence of planetary companions and host metallicity is 
significant at the $\sim 4 \sigma$ level and that the anti-correlation between the presence of stellar companions and 
host metallicity is significant at the $\sim 2 \sigma$ for the $d < 25$ pc FGK sample. Splitting both samples into FG 
and K spectral type stars suggests that the correlation between the presence of planetary companions and host metallicity 
is independent of spectral type but that the anti-correlation between the presence of stellar companions and host metallicity 
is a strong function of spectral type with the anti-correlation disappearing for the bluer FG host stars (see Fig. \ref{fig:slopes}). 
We find no spectral-type dependent binary detection efficiency bias that can explain this anti-correlation.


\vspace{0.5cm}
\section{Is the Anti-Correlation between Metallicity and Stellar Binarity Real?}
\label{sec:binary}

We further examine the relationship between stellar metallicity and binarity by comparing our sample with that of the 
Geneva and Copenhagen survey (GC) of the solar neighbourhood \citep{Nordstrom04} that has been selected as a 
magnitude-limited sample, a volume-limited portion ($d < 40$ pc) of which we analyse. This selection criteria infers 
that the sample is kinematically unbiased, i.e., the sample contains the same proportion of thin, thick and halo stars 
as is found in the solar neighbourhood. We also compare our sample with that of the ``Carney-Latham" survey (CL) that 
has been kinematically selected to have high proper motion stars \citep{Carney87}, i.e., it contains a larger proportion 
of halo stars compared to disk stars than is observed for the solar neighbourhood. 

Our sample is based on the Hipparcos sample that has a limiting magnitude for completeness of $V = 7.9 + 1.1\sin|b|$ 
\citep{Reid02} where $b$ is Galactic latitude. Thus the Hipparcos sample is more complete for stars at higher Galactic 
latitudes where the proportion of halo stars to disk stars increases. Hence our more distant ($25 < d < 50$ pc) 
sample will have a small kinematic bias in that it will have an excess of halo stars, whereas our closer ($d < 25$ pc) 
sample will be less kinematically biased. 


\begin{figure}[!t]
\epsscale{1.1}
\plotone{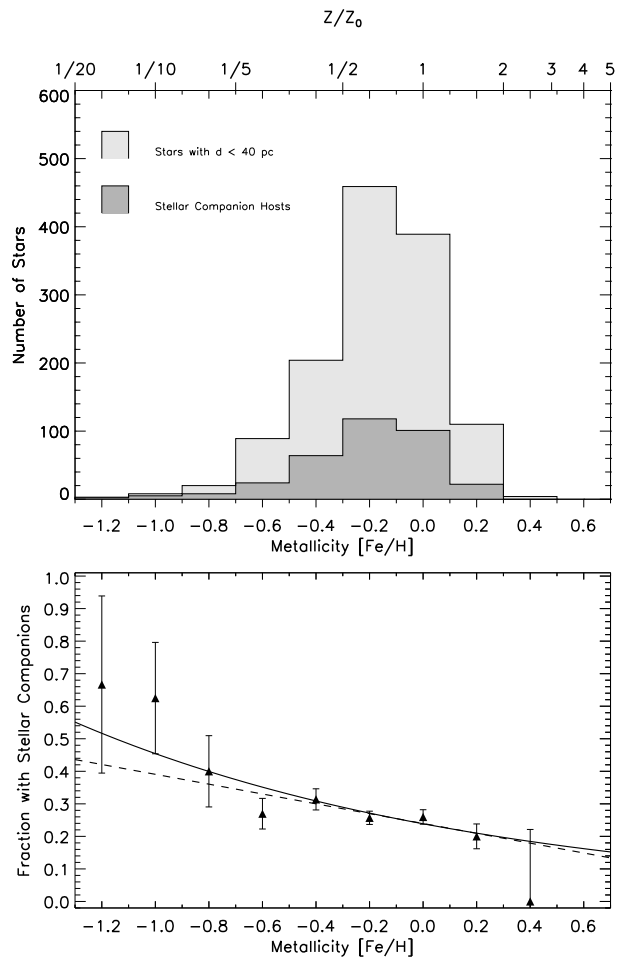}
\caption{Histogram of stars in the complete volume limited GC sample ($d < 40$ pc). We exclude those stars with 
$b-y < 0.3$ so that the spectral type range becomes F7-K3 and thus similar to that of our sample. We only include 
those stars that have between 2 and 10 radial velocity measurements with the CORAVEL spectrograph. We find an 
anti-correlation between binarity and host metallicity as shown by the linear and non-linear best-fits represented 
by the dashed and solid lines respectively.
}
\label{fig:FeH_N04}
\end{figure}

\begin{figure}[!t]
\epsscale{1.1}
\plotone{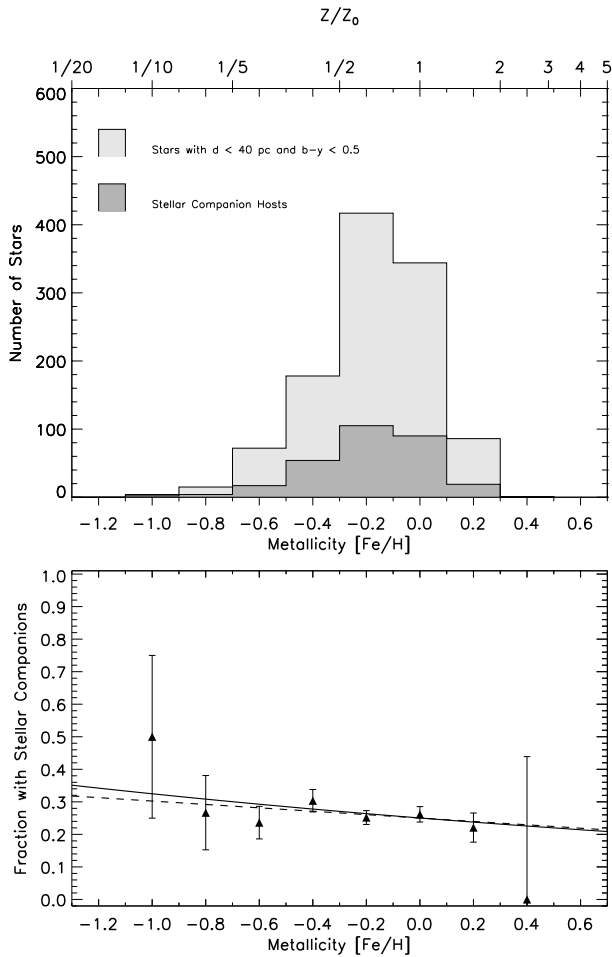}
\caption{Same as Fig. \ref{fig:FeH_N04} but only for the FG dwarfs in the GC sample of stars ($d < 40$ pc). We define 
FG dwarfs as those with $b - y < 0.5$ ($B-V \lsim 0.75$). We find only a marginal anti-correlation between binarity and host 
metallicity as shown by the linear best-fit with gradient $a = -0.05 \pm 0.06$ and the non-linear best-fit with 
$\alpha = -0.11 \pm 0.10$. Using $P_{\rm avg} = 26.1\%$, the scaled linear gradient $a' = (-0.05 \pm 0.06)/26.1\% = -0.2 \pm 0.2$.
}
\label{fig:FeH_N04_FG}
\end{figure}

\begin{figure}[!t]
\epsscale{1.1}
\plotone{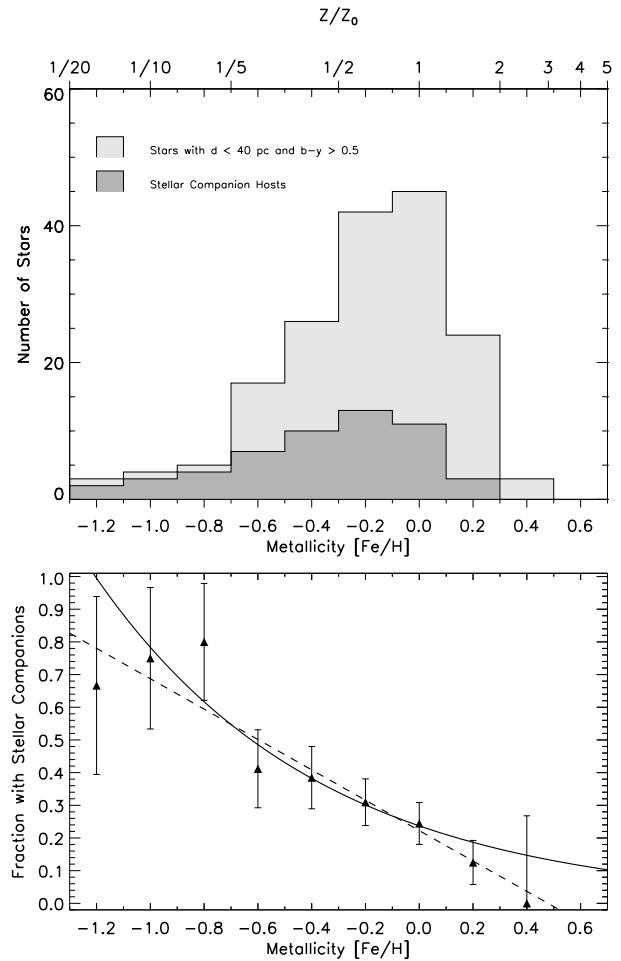}
\caption{Same as Fig. \ref{fig:FeH_N04} but only for the K dwarfs in the GC sample of stars ($d < 40$ pc). We 
define K dwarfs as those with $b - y > 0.5$ ($B-V \gsim 0.75$). We find a very strong anti-correlation between binarity 
and host metallicity as shown by the linear best-fit with gradient $a = -0.46 \pm 0.10$. The non-linear best-fit is 
$\alpha = -0.52 \pm 0.11$. The scaled linear gradient $a' = (-0.46 \pm 0.10)/32.0\% = -1.4 \pm 0.3$. Comparing this plot 
with Fig. \ref{fig:FeH_N04_FG} suggests that the anti-correlation between binarity and host metallicity is stronger 
for redder stars. 
}
\label{fig:FeH_N04_K}
\end{figure}


\subsection{Comparison with a Kinematically Unbiased Sample}

The GC sample contains primarily F and G dwarfs with apparent visual magnitudes $V \lsim 9$ and is complete in volume for 
$d < 40$ pc for F0-K3 spectral type stars. Unlike our sample analyzed in Section \ref{sec:analysis}, it also includes 
early F spectral type stars. The GC sample color range is defined in terms of $b-y$ not $B-V$ like our samples. We remove 
these early F stars with $b-y < 0.3$ \citep[$B-V \lsim 0.5$,][]{Cox00} from the sample so that the GC sample spectral type range 
is similar to ours. The GC sample then ranges from $0.3 \leq b-y \leq 0.6$ ($0.5 \lsim B-V \lsim 1.0$) with those stars 
above $b-y = 0.5$ ($B-V \sim 0.75$) referred to as K stars. We also exclude suspected giants from the GC sample. 

For the GC sample, we only include those binaries observed by CORAVEL between 2 and 10 times so as to avoid a potential 
bias where low metallicity stars were observed more often, thus leading to a higher efficiency for finding binaries around 
these stars. This homogenizes the binary detection efficiency such that any real signal will not be removed by such a 
procedure. Unlike our sample for which we only include binaries with $P < 5$ years, the GC sample also includes much 
longer period visual binaries in addition to short period spectroscopic binaries such that the total binary fraction 
of all types corresponds to $\sim 25\%$. Comparing this with the period distribution for G dwarf stars of 
\citet{Duquennoy91} this binary fraction corresponds to binary systems with periods less than $\sim 10^{5}$ days.

For the volume limited $d < 40$ pc sample we again find an anti-correlation between binarity and stellar host metallicity 
as shown in Fig. \ref{fig:FeH_N04}. Both the linear and non-linear best-fits listed in Table \ref{table:bestfit} are 
significant at or above the $3 \sigma$ level. We also split the GC sample into FG and K spectral type stars in 
Figs. \ref{fig:FeH_N04_FG} and \ref{fig:FeH_N04_K} respectively. The anti-correlation between the presence of stellar 
companions and host metallicity is significant at less than the $1 \sigma$ level for FG stars but significant at the 
$\sim 4 \sigma$ level for K stars. 

These results are qualitatively the same as those found for our sample but quantitatively weaker as shown in Fig. \ref{fig:slopes} 
(confer rows of points labeled GC). This may be due to the higher fraction of late F and early G spectral type stars 
compared to our samples or the larger range ($\sim 10^5$ days) in binary periods contained in the GC sample compared to our 
sample where $P < 5$ years. Another way of interpreting this anti-correlation between binarity and metallicity may be in 
terms of the age and nature of different components of the Galaxy described by stellar kinematics, i.e., F stars are generally 
younger than K stars and thus are more likely to belong to the younger thin disk star population than the older thick disk 
star population. Hence we examine our results in terms of stellar kinematics.


\subsection{Comparison with a Kinematically Biased Sample}

We also compare our samples and that of the GC survey with the \citet{Carney87} high proper motion survey (CL). The CL survey 
contains all of the A, F and early G, many of the late G and some of the early K dwarfs from the Lowell Proper Motion Catalog 
\citep{Giclas71, Giclas78} and which were also contained in the NLTT Catalog \citep{Luyten79, Luyten80}. The number of stars 
in this distribution increases as the stellar colors become redder, peaking at about $B-V = 0.65$, following which the numbers 
of stars begin to decrease \citep{Carney94}. This group has also obtained data for a smaller number of stars from the 
sample of \citet{Ryan89} who sampled sub-dwarfs (metal-poor stars beneath the main-sequence) that have a high fraction of 
halo stars in the range $0.35 < B-V < 1.0$. We refer to this combined sample as outlined in \citet{Carney05} as the CL sample. 
This CL sample contains all binaries detected as spectroscopic binaries, visual binaries or common proper motion pairs. 

\begin{figure}[!t]
\epsscale{1.1}
\plotone{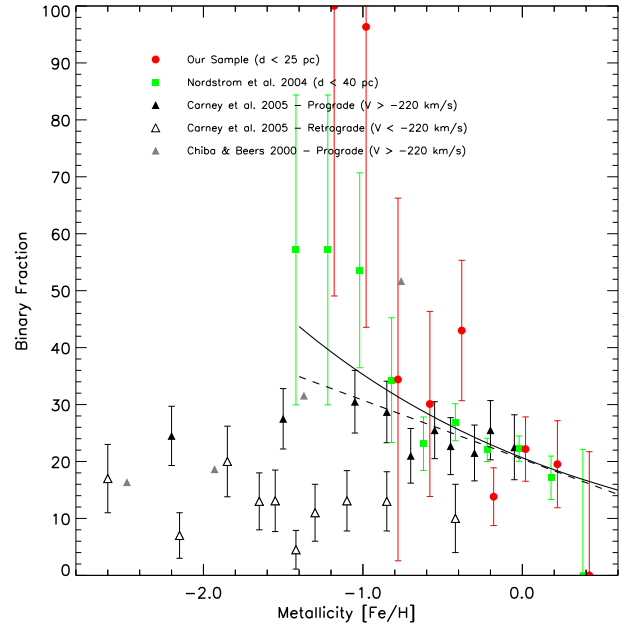}
\caption{This plot is adapted from Fig. 3 of \citet{Carney05}. The black triangles are the points from the CL sample 
of proper motion stars with prograde Galactic tangential velocities. We overplot the binary fraction as a function of 
host metallicity for our close ($d < 25$ pc) F7-K3 sample (Fig. \ref{fig:FeH_Hist}) with red circles and the green 
squares are from the volume limited GC sample ($d < 40$ pc) for F7-K3 stars (Fig. \ref{fig:FeH_N04}). The three samples 
contain different average binary fractions because the period range and the levels of completeness of the stellar 
companions varies between the samples as discussed in the text. We normalise the distributions by scaling our sample 
and the GC sample so that they contain the same fraction of binary stars as the sample of \citet{Carney05} at $[\FeH] = 0$. 
The linear and non-linear best-fits to the three samples combined are shown as dashed and solid lines respectively.
}
\label{fig:Carney_FeH}
\end{figure}

In Fig. \ref{fig:Carney_FeH} we plot the binary fraction of stars on prograde and retrograde Galactic orbits as shown in 
Fig. 3 of \citet{Carney05}. All of the CL stars have $[\FeH] \leq 0.0$. The CL distribution contains a small subset of 
metal-poor $[\FeH] \leq -0.2$ stars from \citet{Ryan89} that has a one-third lower prograde binary fraction due to 
fewer observations. Thus stars with metallicities of between $-0.2$ and $0.0$ have a higher binary fraction than the rest of 
the CL distribution. We make a small correction for this bias in the binary fraction in the range $-0.2 < [\FeH] < 0.0$ 
by lowering the 2 highest metallicity prograde points of the CL distribution by $2\%$. 

We note an anti-correlation between the binary fraction and metallicity for $-1.3 < [\FeH] < 0.0$ range of prograde disk 
stars of the CL distribution as shown in Fig. \ref{fig:Carney_FeH}. We find that the linear best-fit to this anti-correlation
has a gradient of $a = -0.07 \pm 0.06$ and the non-linear best-fit has $\alpha = -0.12 \pm 0.09$, which are both significant 
at slightly above the $1 \sigma$ level. For consistency we exclude the two lowest metallicity points from this best-fit 
so that we analyse the same region of metallicity as our samples and the GC sample and because these two low metallicity 
points will probably contain a significant fraction of halo stars. The average binary fraction is $P_{\rm avg} = 26\%$ for 
the disk-dominated part of the prograde CL distribution. \citet{Carney05} found no correlation between binarity and host 
metallicity for the retrograde halo stars.

We overplot our $d < 25$ pc binary fraction (from Fig. \ref{fig:FeH_Hist}) along with the GC $d < 40$ pc binary fraction 
(from Fig. \ref{fig:FeH_N04}) onto the prograde CL sample in Fig. \ref{fig:Carney_FeH}. All three of these samples have different 
binary period ranges and levels of completeness. We scale our sample and the GC sample to the size of the \citet{Carney05} 
sample by scaling the distributions to contain the same number of binary stars at solar metallicity. The most metal-poor point 
in our close binary distribution is scaled above $100\%$, hence we set this point to $100\%$. The combined three sample 
distribution shows an anti-correlation between binarity and metallicity. The normalized linear best-fit to this is 
$a' = (-0.10 \pm 0.03)/25.7\% = -0.39 \pm 0.12$ and the non-linear best-fit is $\alpha = -0.22 \pm 0.05$ which are both 
significant at or above the $\sim 3 \sigma$ level (see last row of Table \ref{table:bestfit}). This combined result is our 
best estimate and indicates a strong anti-correlation between stellar companions and metallicity for $[\FeH] > -1.3$.


\begin{figure}[!t]
\epsscale{1.1}
\plotone{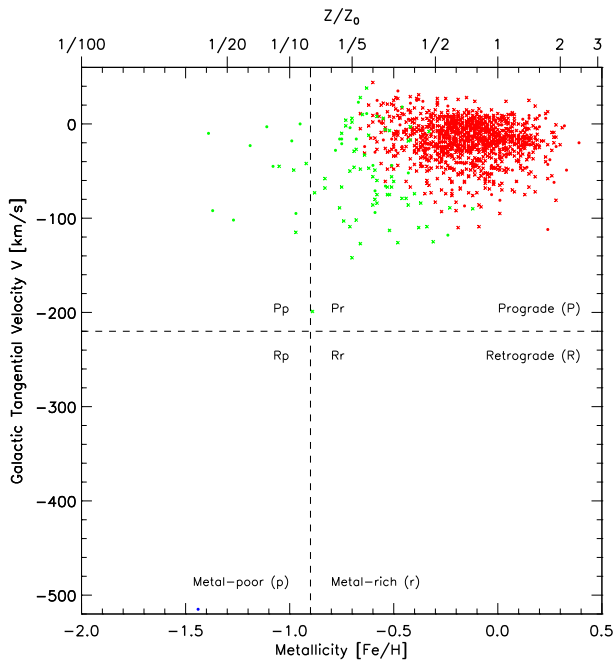}
\caption{We plot tangential Galactic velocity $V$ as a function of metallicity $[\FeH]$ for the kinematically unbiased 
GC sample ($d < 40$ pc). We use a probabilistic method to assign the stars in the GC sample to the three Galactic 
populations (halo, thick and thin disks) as discussed in the Appendix. Red points are thin disk stars, green points are 
thick disk stars and a blue point is the single halo star in the sample at $V < -500$ km/s. Crosses represent FG 
spectral-type stars and circles K stars. The ratio of thick/thin disk stars is $\sim 3$ times higher for K stars 
than for FG stars.
}
\label{fig:V_FeH}
\end{figure}

\begin{figure}[!t]
\epsscale{1.1}
\plotone{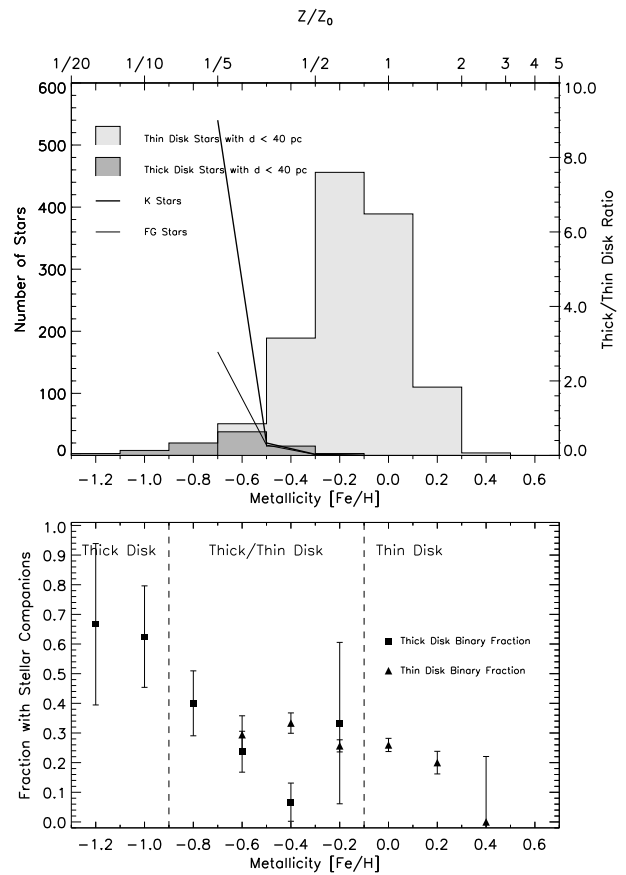}
\caption{Histogram of the stars in Fig. \ref{fig:V_FeH} suspected of belonging to the thick disk and the thin disk in 
the GC sample ($d < 40$ pc). Notice the difference by a factor $\sim 2$ between the higher binary fraction thick disk 
stars and the lower binary fraction thin disk stars. We note that the K star distribution contains a higher ratio of 
thick disk stars than the FG star distribution in the thick/thin disk overlap region.
}
\label{fig:disc}
\end{figure}


\subsection{Discussion}

We examine our results in terms of Galactic populations by determining the most likely population membership (halo, thick 
or thin disk) for each star in the GC sample using the method outlined in the Appendix and then plotting them in the 
Galactic tangential velocity $V$ - metallicity $[\FeH]$ plane as in Fig. \ref{fig:V_FeH}. We use red points for the thin 
disk stars, green points for the thick disk stars and a blue point for the single halo star. The kinematically unbiased 
GC sample contains mostly thin disk stars. Excluding the one halo star in the GC sample, stars with $[\FeH] \lsim -0.9$ 
belong to the thick disk and stars with $[\FeH] \gsim -0.1$ belong to the thin disk. The region $-0.9 \gsim [\FeH] \gsim -0.1$ 
contains a combination of both thick and thin disk stars. 

We also plot these thick and thin disk stars as separate histograms in metallicity in Fig. \ref{fig:disc}. In the region 
$[\FeH] \lsim -0.9$ that contains only thick disk stars, we find that the binary fraction is approximately twice as large 
as for the region $[\FeH] \gsim -0.1$ that contains only thin disk stars. In both of these single population regions the 
binary fraction also appears to be approximately independent of metallicity. While our purely probabilistic method of assigning 
the stars in the GC sample to Galactic populations is useful for determining the general regions of parameter space that the 
individual populations occupy, it is not precise enough to show exactly which stars belong to which population. This is 
especially true for the regions of parameter space that have large overlaps such as that between the thick and thin disk stars 
in Fig. \ref{fig:V_FeH}. Thus the thin and thick disk binary fractions in the interval $-0.9 \lsim [\FeH] \lsim -0.1$ are 
probably mixtures. We suspect that the thin and thick disk binary fractions in this overlap region will remain at the 
same levels as found for the non-overlapping regions. The anti-correlation between binarity and metallicity in the 
$-0.9 \lsim [\FeH] \lsim -0.1$ range may be due to this overlap between higher binarity thick disk stars and lower binarity 
thin disk stars.

We now partition the Galactic tangential velocity $V$ - metallicity $[\FeH]$ parameter space into four quadrants. We split 
the $V$ parameter space into those stars on prograde Galactic orbits (P) and those on retrograde Galactic orbits (R). We 
split the $[\FeH]$ parameter space into those stars that are metal rich (r) with $[\FeH] \gsim -0.9$ and those that are 
metal poor (p) with $[\FeH] \lsim -0.9$. We then label these quadrants by the direction of Galactic orbital motion followed 
by the range in metallicity or Pp, Pr, Rr and Rp as shown in Fig. \ref{fig:V_FeH}. We now assume that the Pp quadrant contains 
a mixture of halo and thick disk stars and that the Pr quadrant contains a mixture of thin and thick disk stars and that the 
Rp and Rr quadrants only contain halo stars. 

The combined anti-correlation between binarity and metallicity in Fig. \ref{fig:Carney_FeH}, that all three samples appear to 
have in common, is predominantly in the Pr quadrant of $V - [\FeH]$ parameter space that contains a mixture of thick and thin 
disk stars. As discussed above this anti-correlation may be due to the overlap of high binarity thick disk stars and lower 
binarity thin disk stars.

While \citet{Latham02} suggest that the halo and disk populations have the same binary fraction, \citet{Carney05} find lower
binarity in retrograde stars. As shown in Fig. \ref{fig:Carney_FeH} there is a clear difference of about a factor of 2 in the 
region $[\FeH] \gsim -0.9$ between the binary fractions of prograde disk stars compared to retrograde halo stars (Pr and Rr 
respectively). All the retrograde halo stars appear to have the same binary fraction (quadrants Rr and Rp). The Pp quadrant 
contains prograde halo stars and has a $\sim 2$ times higher binary fraction than the quadrants containing retrograde halo 
stars. However the Pp quadrant also contains thick disk stars in addition to prograde halo stars.

We propose that the Pp region, $[\FeH] \lsim -0.9$, for stars on prograde Galactic orbits contains a mixture of low binarity 
halo stars and high binarity thick disk stars. In Fig. \ref{fig:Carney_FeH} at $[\FeH] \sim -0.9$ our close sample ($d < 25$ pc) 
and the GC sample ($d < 40$ pc) start to diverge from the data points of the CL survey. This observed divergence may be due to 
the CL survey being comprised of high proper motion stars and consequently a higher fraction of prograde halo stars compared 
to thick disk stars than the kinematically unbiased GC sample and our relatively kinematically unbiased sample where thick disk 
stars probably numerically dominate over halo stars. 

Using a kinemically unbiased sample, \citet{Chiba00} report that for the three regions $-1.0 > [\FeH] > -1.7$, $-1.7 > [\FeH] > -2.2$ 
and $-2.2 > [\FeH]$ that the fraction of stars that belong to the thick disk are $29\%$, $8\%$ and $5\%$ respectively, with the 
rest belonging to the halo. We restrict these thick disk fraction estimates to stars only on prograde orbits by assuming that 
all of the thick disk stars are on prograde orbits and that half of the halo stars are prograde and the other half are retrograde. 
Thus the fraction of prograde stars that are thick disk stars is $45\%$, $15\%$ and $10\%$ for the three metallicity regions 
respectively.

Using these three prograde restricted thick disk/halo ratios reported in \citet{Chiba00} combined with observed binary fraction 
for the thick disk (55\%) and halo (12\%) stars in Fig. \ref{fig:Carney_FeH} we can test the proposal in the Pp quadrant that 
the two lowest metallicity prograde points from \citet{Carney05} contain a mixture of low binarity halo stars and high binarity 
thick disk stars. We plot the three estimated mixed thick disk/halo binary fraction points as grey triangles in Fig. \ref{fig:Carney_FeH}. 
We note that they are consistent with the two prograde \citet{Carney05} points thus supporting our proposal that the Pp quadrant 
contains a mixture of low binarity halo stars and high binarity thick disk stars. These mixed thick disk/halo points also 
show a correlation between the presence of stellar companions and metallicity for stars in the Pp region. 

Our results suggest that thick disk stars have a higher binary fraction than thin disk stars which in turn have a higher 
binary fraction than halo stars. Thus for stars on prograde Galactic orbits we observe an anti-correlation between binarity 
and metallicity for the region of metallicity $[\FeH] \gsim -0.9$ that contains an overlap between the lower-binarity, 
higher-metallicity thin disk stars and the higher-binarity, lower-metallicity thick disk stars. We also find for stars on 
prograde Galactic orbits, a correlation between binarity and metallicity for the range $[\FeH] \lsim -0.9$, that contains 
an overlap between the higher-binarity, higher-metallicity thick disk stars and the lower-binarity, lower-metallicity halo 
stars.


\section{Summary}

We examine the relationship between Sun-like (FGK dwarfs) host metallicity and the frequency of close companions 
(orbital period $< 5$ years). We find a correlation at the $\sim 4 \sigma$ level between host metallicity and the presence 
of planetary companion and an anti-correlation at the $\sim 2 \sigma$ level between host metallicity and the presence of 
a stellar companion. We find that the non-linear best-fit is $\alpha = 2.2 \pm 0.5$ and $\alpha = -0.8 \pm 0.4$ for 
planetary and stellar companions respectively (see Table \ref{table:bestfit}).

\citet{Fischer05} also quantify the planet metallicity correlation by fitting an exponential to a histogram in $[\FeH]$. 
They find a best-fit of $\alpha = 2.0$. Our result of $\alpha = 2.2 \pm 0.3$ is a slightly more positive correlation and 
is consistent with theirs. Our estimate is based on the average of the best-fits to the metallicity data binned both as 
a function of $[\FeH]$ and $\Z/\Z_{\sun}$. Larger bins tend to smooth out the steep turn up at high $[\FeH]$ and may be 
responsible for their estimate being slightly lower.

We also analyze the sample of \citet{Nordstrom04} and again find an anti-correlation between metallicity and close stellar 
companions for this larger period range. We also find that K dwarf host stars have a stronger anti-correlation between 
host metallicity and binarity than FG dwarf stars. 

We compare our analysis with that of \citet{Carney05} and find an alternative explanation for their reported binary frequency 
dichotomy between stars on prograde Galactic orbits with $[\FeH] \lsim 0$ compared to stars on retrograde Galactic orbits with 
$[\FeH] \lsim 0$. We propose that the region, $[\FeH] \lsim -0.9$, for stars on prograde Galactic orbits contains a mixture 
of low binarity halo stars and high binarity thick disk stars. Thick disk stars appear to have a $\sim 2$ higher binary fraction 
compared to thin disk stars, which in turn have a $\sim 2$ higher binary fraction than halo stars. 

While the ratio of thick/thin disk stars is $\sim 3$ times higher for K stars than for FG stars we only observe a marginal 
difference in their distributions as a function of metallicity. In the region $-0.9 \lsim [\FeH] \lsim -0.1$ that we suspect 
contains a mixture of thick and thin disk stars, the K star distribution contains a higher ratio of thick disk stars 
compared to the FG star distribution at a given metallicity. This difference is marginal but can partially explain the 
kinematic and spectral-type ($\sim$ mass) results. 

Thus for stars on prograde Galactic orbits as we move from low metallicity to high metallicity we move through low binarity 
halo stars to high binarity thick disk stars to medium binarity thin disk stars. Since halo, thick disk and thin disk stars 
are not discrete populations in metallicity and contain considerable overlap, as we go from low metallicity to high metallicity 
for prograde stars, we firstly observe a correlation between binarity and metallicity for the overlapping halo and thick 
disk stars and then an anti-correlation between binarity and metallicity for the overlapping thick and thin disk stars.  


\acknowledgements

We would like to thank Johan Holmberg for his help on analysing the Geneva sample and Chris Flynn, John Norris, 
Virginia Trimble, Richard Larson, Pavel Kroupa and David Latham for helpful discussions. 


\clearpage
\appendix
\section{Probability of Galactic Population Membership}
\label{sec:appendix}

We use a similar method to that of \citet{Reddy06} in assigning a probability to each star of being a member of the thin disk, 
thick disk or halo populations. We assume the GC sample is a mixture of the three populations. These populations are assumed 
to be represented by a Gaussian distribution for each of the 3 Galactic velocities $U,V,W$ and for the metallicity $[\FeH]$. 
The age dependence of the quantities for the thin disk are ignored. The equations establishing the probability that a star 
belongs to the thin disk ($P_{\rm thin}$), the thick disk ($P_{\rm thick}$) or the halo ($P_{\rm halo}$) are

\begin{equation}
P_{\rm thin} = f_1 \frac{P_1}{P} \; , \; P_{\rm thick} = f_2 \frac{P_2}{P} \; , \; P_{\rm halo} = f_3 \frac{P_3}{P}
\end{equation} 

where 

\begin{eqnarray}
P     &=& \sum f_{i} P_{i} \\
P_{i} &=& C_{i} \exp \left[ -\frac{U^2}{2\sigma_{U_{i}}^2} -\frac{(V- \langle V\rangle)^2}{2\sigma_{V_{i}}^2} 
          -\frac{W^2}{2\sigma_{W_{i}}^2} -\frac{([\FeH]- \langle [\FeH]\rangle)^2}{2\sigma_{[\FeH]_{i}}^2} \right] \nonumber \\
C_{i} &=& \frac{1}{\sigma_{U_{i}} \sigma_{V_{i}} \sigma_{W_{i}} \sigma_{[\FeH]_{i}}} (i = 1,2,3) \nonumber
\end{eqnarray}

\noindent
Using the data in Table \ref{table:pop} taken from \citet{Robin03} we compute the probabilities for stars in the GC sample. 
For each star, we assign it to the population (thin disk, thick disk or halo) that has the highest probability. We plot the 
probable halo, thick and thin disk stars of the GC sample in Fig. \ref{fig:V_FeH}.


\begin{deluxetable}{cccccccc}
\tablecaption{Properties of the Three Stellar Populations\label{table:pop}}
\tablehead{\colhead{Component} & \colhead{$\sigma_{U}$} & \colhead{$\langle V\rangle$} & \colhead{$\sigma_{V}$} & \colhead{$\sigma_{W}$}
& \colhead{$\langle [\FeH]\rangle$} & \colhead{$\sigma_{[\FeH]}$} & \colhead{Fraction $f$}}
\startdata
Thin Disc  & 43  & -15  & 28  & 17 & -0.1 & 0.2 & 0.925 \\
Thick Disc & 67  & -53  & 51  & 42 & -0.8 & 0.3 & 0.070 \\
Halo       & 131 & -226 & 106 & 85 & -1.8 & 0.5 & 0.005 \\
\enddata
\tablecomments{Data from \citet{Robin03}.}
\end{deluxetable} 



\end{document}